\documentclass[sigchi]{acmart}%
\definecolor{RED}{HTML}{ff2e63}
\definecolor{TURK}{HTML}{08D9D6}
\definecolor{BLUE}{HTML}{0000ff}
\usepackage[utf8]{inputenc}%
\usepackage{longtable}%
\usepackage{siunitx}%
\usepackage{array}%
\usepackage{tikz}%
\usetikzlibrary{decorations.pathreplacing}%
\usepackage{tabularx}%
\usepackage{color}%
\newcommand\transition[1]{{\color{black}{#1}}}%
\usepackage{multirow}%
\usepackage{makecell}%
\usepackage{caption}%
\usepackage{subcaption}%
\usepackage{tikz}%
\usetikzlibrary{backgrounds,positioning,shapes.geometric, arrows}%
\copyrightyear{2025}
\acmYear{2025}
\setcopyright{cc}
\acmConference[CHI '25]{CHI Conference on Human Factors in Computing Systems}{April 26-May 1, 2025}{Yokohama, Japan}
\acmBooktitle{CHI Conference on Human Factors in Computing Systems (CHI '25), April 26-May 1, 2025, Yokohama, Japan}\acmDOI{10.1145/3706598.3713464}
\acmISBN{979-8-4007-1394-1/25/04}
\newcommand{\NUMARTICLES}{11,542}%
\begin{document}%

\title[Keeping Score]{%
Keeping Score: 
A Quantitative Analysis
of 
How the CHI Community Appreciates
Its Milestones
}%

\author{Jonas Oppenlaender}%
\email{jonas.oppenlaender@oulu.fi}%
\orcid{0000-0002-2342-1540}%
\affiliation{%
  \institution{University of Oulu}%
  \city{Oulu}%
  \country{Finland}%
}

\author{Simo Hosio}%
\email{simo.hosio@oulu.fi}%
\orcid{0000-0002-9609-0965}%
\affiliation{%
  \institution{University of Oulu}%
  \city{Oulu}%
  \country{Finland}%
}

\begin{abstract}%
The ACM CHI Conference has a tradition of citing its intellectual heritage.
At the same time, we know CHI is highly diverse and evolving.
In this highly dynamic context, it is not clear how the CHI community continues to
    appreciate
its milestones (within and outside of CHI).
We present an investigation into how the community's citations to milestones have evolved over 43~years of CHI Proceedings (1981--2024).
Forgetting curves plotted for each year 
suggest that milestones are slowly fading from the CHI community's collective memory.
However, the picture is more nuanced when we trace citations to the top-cited milestones over time.
We identify three distinct types of milestones cited at CHI,
a typology of milestone contributions,
and define the Milestone Coefficient as a metric to assess the impact of milestone papers on a continuous scale.
Further, we provide empirical evidence of a Matthew effect at CHI.
We discuss the broader 
ramifications for the CHI community and the field of HCI.
\end{abstract}%


%
%
\begin{teaserfigure}%
\centering
  \includegraphics[width=\textwidth]{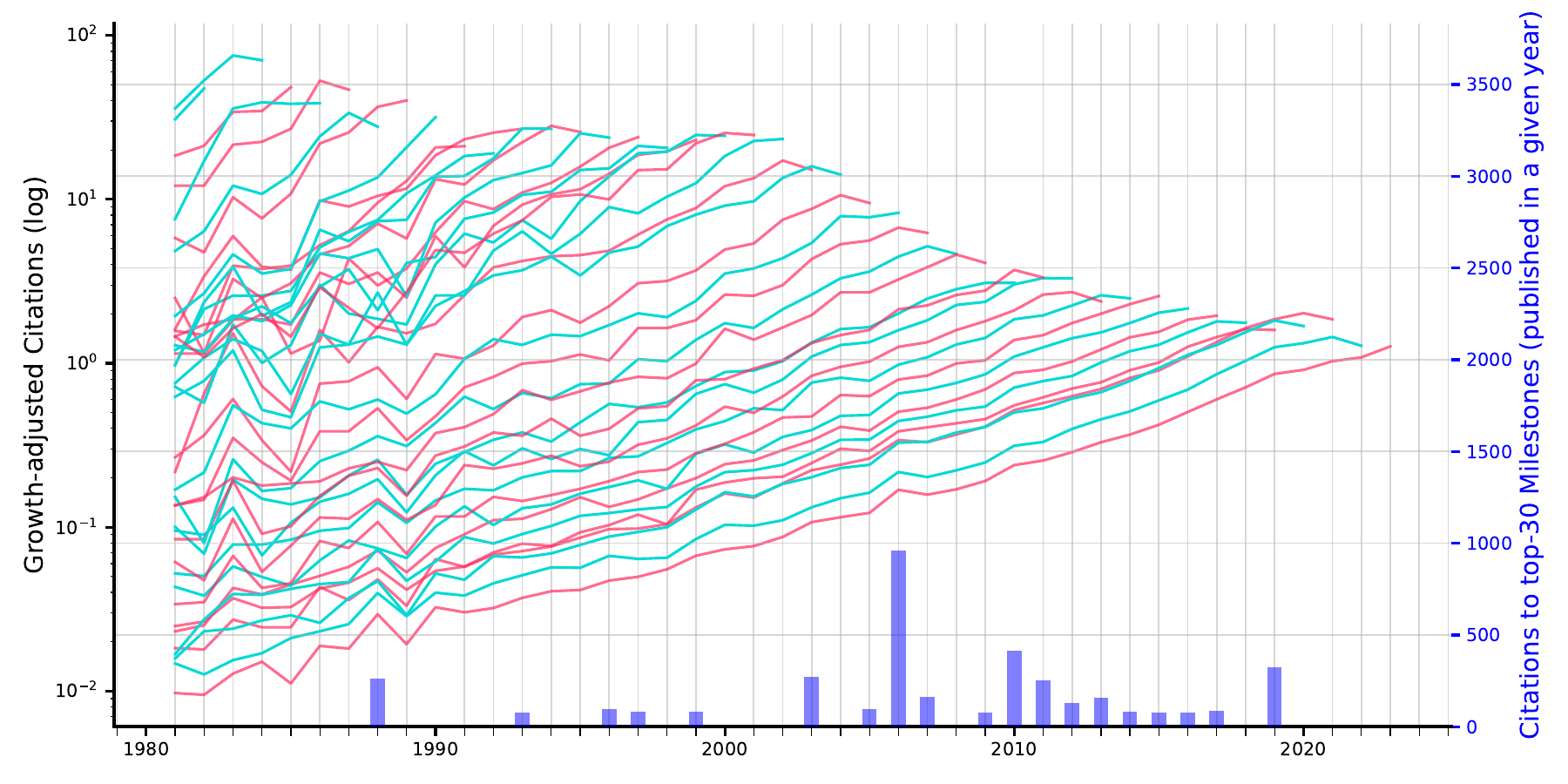}%
  \caption{%
  Forgetting curves~\cite{Reisz_2022_New_J._Phys._24_053041.pdf} for the Proceedings of the ACM Conference on Human Factors in Computing Systems (CHI) from 1981 to 2024.
A forgetting curve plots the number of citations from CHI papers to prior papers published during the years plotted on the x-axis.
The primary y-axis represents the 
citations in these publications over time, adjusted for
overall growth in the number of publications in the proceedings, and scaled logarithmically for comparison.
      Each line represents the forgetting curve for papers published in the year following the right-most point of the line.
          For instance, the curve at the bottom is the forgetting curve for the 2024 CHI proceedings.
  The peaks in the forgetting curves visually indicate the presence of ``milestones'', when read along a vertical line from top to bottom.
  In other words, several proceedings contain particularly many references to milestones published in that year. 
  For forgotten milestones, this visible peak is lost over time.
    Here, `forgetting' refers to a decrease in citations over time.
    We can see how the proceedings of the last few years are distributing their citations more uniformly, with less spikes and dips visible in the curves.
    This is indicative of the community's divided attention outside what was in earlier years considered as its key milestones.
  Independent of the forgetting curves, the secondary y-axis depicts the number of citations from all CHI papers ever published to the top-30 milestones most cited at CHI (see \autoref{tab:milestones}) per their publication years (x-axis).
        For instance, \citeauthor{M1}'s milestone in the year 2006 \cite{M1}
        received many citations and the adoption of this milestone in the CHI community is 
        visible as a growing peak in the forgetting curves.
  The fact that there were no proceedings in 1984 contributes to fluctuations in the older curves.
  }%
  \label{fig:teaser}%
\end{teaserfigure}%
\maketitle%
%
%
\section{Introduction}%
\label{sec:introduction}%

Human-Computer Interaction (HCI) is highly interdisciplinary
and 
characterized by constant evolution and changes 
\cite{10.1145/3477107,2804405.pdf}.
    Recent changes include, to name but a few,
    an overall growth in the number of scientific papers published each year
    \cite{science.aao0185.pdf,s41599-021-00903-w.pdf}
    and an acceleration of pace in science with mounting career and publication pressures~\cite{edwards-roy-2017-academic-research-in-the-21st-century-maintaining-scientific-integrity-in-a-climate-of-perverse.pdf,oppenlaender-citations-HCI}. In addition, sudden technological advances can shift focus and topics in HCI research.
One such technological advance is the recent progress in 
large auto-regressive language models, such as OpenAI's ChatGPT \cite{bubeck2023sparksartificialgeneralintelligence,NIPS2017_3f5ee243}. 
The increasing integration of language models into HCI research reflects broader trends in methodological acceleration and automation \cite{3699689.pdf,pang2025understandingllmificationchiunpacking}.
While novel AI tools can streamline certain tasks, over-reliance on such tools and their misuse has led to concerns about prompt-hacking (akin to p-hacking)~\cite{3673861.pdf} and shallow inquiries that prioritize speed over depth \cite{3699689.pdf,2406.07016.pdf,2403.16887.pdf,2403.07183.pdf,SuspectedAI,llmwrappers}.
As HCI evolves, it is critical to reflect on the types of contributions that drive long-lasting impact, particularly amid shifting research norms and topics.
In light of this dynamic context, it is important to explore what we value in HCI and to understand how the CHI community interacts with its intellectual heritage.



The CHI community has a history of citing and valuing its intellectual heritage \cite{oppenlaender-citations-HCI}.
For instance, the ``Guide to a Successful Paper'' at CHI '16
expressively called authors to \textit{``make sure to cite prior work [...] in the relevant area''}
    and works 
    \textit{``that have had a major influence on your own work''} \cite{chi2016-guide}.
Letting others build on one's work has been quoted to be \textit{``the entire purpose of a CHI Paper''} and \textit{``lack of references''} was identified as \textit{``a frequent cause for complaint -- and low rating -- by reviewers''} \cite{chi2016-guide}.
These sentiments help steer CHI, highlighting the
community's commitment
to acknowledging relevant prior work.
The emphasis on recognizing 
intellectual heritage highlights the importance of understanding which works have shaped the field of HCI. However, beyond subjective appreciation, a more systematic approach is needed to quantify the influence of these impactful works.

Citation analysis, while not the only one, is the de facto means for analyzing the impact of past contributions. In this work, we explore quantitatively how the CHI community values its past \textit{milestones}: articles that have attained exceptional attention from the community. These articles are important to the community, and thus to the discipline itself. For instance,
    the System Usability Scale (SUS)~\cite{SUS} and the NASA-TLX~\cite{M2} are two widely adopted measurement scales used in HCI studies. 
Given the interdisciplinarity of the field, these milestones do not always originate within the community. 
Thematic Analysis guidelines~\cite{M1} are a great example of an influential contribution that did not originate from CHI or even the broader HCI community.
We explore how the CHI community pays attention to its milestones over time, exploring particularly the following research questions:%
\newcommand{\RQOne}{What are the milestones most cited in the CHI community?}
\newcommand{\RQTwo}{What are the contributions of these milestones?}
\newcommand{\RQThree}{How has the CHI community's attention to its milestones evolved over time?
}
\newcommand{\RQFour}{Does the CHI community show a preference for its milestone authors?}
\begin{itemize}%
    \item[RQ1:]%
        \textit{\RQOne}%
    \item[RQ2:]%
        \textit{\RQTwo}%
    \item[RQ3:]%
        \textit{\RQThree}%
    \item[RQ4:]%
        \textit{\RQFour}%
\end{itemize}%

We present the results of a quantitative analysis of \NUMARTICLES~articles published in the Proceedings of the ACM CHI Conference on Human Factors in Computing Systems (1981--2024, with the exception of 1984).
Our \textit{citation curve} analysis examines the CHI community's current attention span and the community's selective memory.
The majority of milestone citations point 5--15 years back in time, suggesting a ``sweet spot'' of relevance for papers published at CHI.
As expected, we also find that the community milestones make diverse types of contributions. Yet, a comparison with existing well-known research contribution taxonomies \cite{2907069.pdf,chi2024}, reveals that popular CHI contribution types (such as prototypes and systems), but also other contribution types (such as surveys, datasets, and replications), are absent in the CHI community's milestones. This highlights how some highly popular contribution types---such as prototypes and systems---all play a role but seldom leave a long-lasting historical mark.
We present the 30~milestones most cited by the CHI community in this paper. In the spirit of Open Science, we share a dataset of the 300~most cited milestones in the Supplemental Material of this article.
%
We further present a case study on CHI's ``super milestone'', 
\citeauthor{M1}'s Thematic Analysis \cite{M1}, that continues to enjoy extraordinary popularity at CHI.
Finally, we present the \textit{Milestone Coefficient} as
a metric for assessing research impact of milestones.
This metric offers a more nuanced understanding of what constitutes a milestone beyond mere citation count. 
We discuss our findings in light of related literature and provide our reflections on how the community remembers, forgets, and curates past knowledge.

Our research contributes to meta-research on HCI literature (``meta-HCI'') \cite{meta-HCI},
particularly focusing on the CHI community, which, while not fully representative, is an important community in the field of~Human-Computer Interaction.

\section{Related Work}%
\label{sec:relatedwork}%

While the field of HCI is buzzing with activity, as evident in the well-visited and growing annual CHI Conference, surprisingly little research and self-reflection on HCI research is being published at CHI. For example, the HCI field's top conference currently has no subcommittee for meta-scientific investigations.
This 
contrasts with the field's critical and self-reflective tradition~\cite{REVD,SCHON}.
Such self-reflective and meta-scientific contributions are critically important to advance the HCI field as a whole, 
and in the following we present prior meta-research on HCI and CHI.%
%
%
%
\subsection{Bibliometric Analyses at CHI}%
\label{sec:rel:citation-analysis}%
%
%
\citeauthor{3290607.3310429.pdf} \cite{3290607.3310429.pdf}
conducted a quantitative analysis
of 6,578 CHI papers, investigating how authors write their papers and how factors such as readability and name dropping influence citation counts.
\citeauthor{1518701.1518810.pdf} \cite{1518701.1518810.pdf} presented a scientometric analysis of CHI proceedings, focusing on organizations and countries that contribute to CHI.
    Their work highlighted the difficulty of judging quality in the context of best paper awards, finding a mismatch between awarded papers and citation counts.
Another bibliometric analysis was presented by \citeauthor{CHI_Bibliometrics_20190301.pdf}\cite{CHI_Bibliometrics_20190301.pdf}.
Their citation network analysis and cluster analysis identified emerging research themes in HCI.
Research themes were also the topic of \citeauthor{oppenlaender2023mapping}'s work on research challenges in HCI~\cite{oppenlaender2023mapping}. The authors analyzed CHI papers with the use of language models and mapped out the diverse research landscape of CHI.
A more traditional 
approach to analyzing topics in CHI was taken by
\citeauthor{2556288.2556969.pdf} who presented a co-word analysis of the evolution of the HCI field \cite{2556288.2556969.pdf}. Their thematic analysis of keywords in CHI papers published between 1994 and 2013 found a fragmentation of research approaches within HCI.
Finally,
\citeauthor{1520340.1520364.pdf} \cite{1520340.1520364.pdf}
investigated repeat authorship at CHI.
Among other findings, \citeauthor{1520340.1520364.pdf} found the mean number of authors in CHI papers is increasing, and many CHI authors are repeat authors.

It can be noted that the above approaches did neither consider the time dimension of citation networks nor preferential attachment \cite{67.pdf,merton1968.pdf}, a self-reinforcing mechanism where successful papers
become more successful over time.
    These factors concern the dynamic nature of scholarly publishing, where older papers have more time to accrue an interest in the community.
        Therefore, a citation analysis must always consider the growth in the number of publications 
        \cite{Reisz_2022_New_J._Phys._24_053041.pdf}.
%
%
%
%
%
%
%
%
%
\transition{%
These dynamics and mechanisms play a key role
in our analysis of the CHI community's milestones.%
}%
%
%
\subsection{Milestones}%
\label{sec:rel:milestones}%
Human-Computer Interaction 
is highly interdisciplinary, borrowing methods and approaches from diverse fields such as Design, Psychology, and Computer Science.
Perhaps it is for this interdisciplinarity
and HCI being an ``inter-discipline''~\cite{2702613.2732505.pdf}
that
milestones have, so far, not been a topic in the HCI literature.
Milestones are highly cited papers that move a field forward.
    In many cases, these papers invented new methods, frameworks, or entire sub-fields in HCI that have inspired hundreds of research papers.
    We feel strongly that milestone is the appropriate name for such papers.
    Moreover, the term milestone is also used in other fields in a similar fashion.
Being highly cited over years, milestone papers
show resilience against aging.
    In this context, aging of a publication refers to the preference of authors to cite more recent works over older works~\cite{Reisz_2022_New_J._Phys._24_053041.pdf,science.1237825,1704.04657.pdf}.

However, there is no standard definition of what constitutes a milestone paper.
Each scientific field moves at a different pace, and even within a given scientific field or discipline, there is no clear definition of what constitutes a milestone.
This is evident in the different approaches taken by authors in prior research to identify milestone papers.
    \citeauthor{49_1_online.pdf} used the total number of citations over a paper's lifespan as criteria, focusing on papers with more than 1,000~citations \cite{49_1_online.pdf}.
    On the other hand, \citeauthor{1310.8220.pdf} examined the 50~best-performing papers \cite{1310.8220.pdf}, and
    \citeauthor{Reisz_2022_New_J._Phys._24_053041.pdf}  focused on the top-30 most-cited papers \cite{Reisz_2022_New_J._Phys._24_053041.pdf}.
    \citeauthor{ke-et-al-2015-defining-and-identifying-sleeping-beauties-in-science.pdf} took an entirely different approach, focusing on wake-up intensity of would-be milestone papers~(``sleeping beauties'') \cite{ke-et-al-2015-defining-and-identifying-sleeping-beauties-in-science.pdf}.
Common to these approaches is that a milestone paper is a work that receives an exceptional number of citations, compared to other papers in a given research field.
\transition{%
In the following section, we review related work on 
a research community's attachment to its milestones.%
}%
%
%
%
%
%
\subsection{Citation Memory}%
\label{sec:rel:citation-memory}%
A research community's collective ``memory'' is expressed through its citations to past works.
Preferential attachment~\cite{67.pdf} is the phenomenon that highly cited papers are more visible and, thus, more likely to be cited than less-cited papers~\cite{67.pdf,science.1237825}.
The preferential attachment of a research community to its high-performing papers (and authors) is also referred to as accumulated advantage or Matthew effect in science~\cite{merton1968.pdf}.


On the other hand, past works may fall out of favor due to factors such as technological progress and shifts in research topics over time.
In these cases, the research community's collective memory may be fading.
In \citeyear{asi.5090110105}, \citeauthor{asi.5090110105}
coined the term half-life of academic literature to describe this phenomenon \cite{asi.5090110105}.
A related concept is citation amnesia, which refers to how far back in time researchers tend to cite 
\cite{2023.acl-long.341v2.pdf}.

It can be concluded that the number of citations a paper receives depends both on the paper’s accrued citations and its age \cite{science.1237825,1704.04657.pdf}.
Ultimately, the absolute number of citations is not the only imaginable indicator for a paper's impact~\cite{1310.8220.pdf},
and citation analysis should always consider the dynamics and exponential growth of the number of publications \cite{Reisz_2022_New_J._Phys._24_053041.pdf,science.1237825,1704.04657.pdf}.
\transition{%
In the following section, we describe how forgetting curves are a means of analysis that considers this growth in publications.%
}%
%
%
%
\subsection{Forgetting Curves}%
\label{sec:rel:forgettingcurves}%

A \textit{citation curve} plots the number of citations to past papers from papers published in a given year.
    For instance, if a paper published in the year 2024 cites a paper published in 1950, this counts as one citation to the year 1950.
Citation curves provide a visual indication of the attention span of authors. On the one hand, a flat citation curve indicates a long community memory where many old papers are being cited in a given proceedings year. On the other hand, a citation curve with a steep and right-leaning peak indicates a short community memory where old papers are quickly being forgotten.
However, one factor that comes into play is the number of articles published each year. This number has been increasing in recent years, thereby introducing a bias to the insights that we can derive from citation curves.


\textit{Forgetting curves} \cite{Reisz_2022_New_J._Phys._24_053041.pdf} address this problem.
A forgetting curve is a growth-adjusted citation curve.
    For all references in a given proceedings year, the forgetting curve looks at what prior years have been cited and counts citations to these years (i.e., the citation curve), adjusted for growth in publications.
The forgetting curve is calculated as follows:\begin{equation*}
\bar{c}_{t_{i} t_{j}} = \frac{N_{t_{j}}}{N} \sum_{t_j} c_{t_{i} t_{j}}
\end{equation*}
where
$N_{t_{j}}$ is the number of papers  published in year $t_j$,
$N$ is the total number of papers published in all years,
and 
$c_{t_{i} t_{j}} $
is the citation curve \cite{Reisz_2022_New_J._Phys._24_053041.pdf}.
The sum $\sum_{t_j} c_{t_{i} t_{j}}$ is the total number of citations from papers published in year $i$ to papers published in year $j$.
    To mitigate bias due to growth in publications over the years, the forgetting curve is 
    weighted with the relative number of publications.
    This weighting factor 
    together with the citation counts can be interpreted as an expectancy value. Valleys in the forgetting curve will form when citations to a past year under-perform  compared to the expected value, and peaks will indicate above average performance.
See the work by \citeauthor{Reisz_2022_New_J._Phys._24_053041.pdf} \cite{Reisz_2022_New_J._Phys._24_053041.pdf} for a more detailed explanation of citation curves and forgetting curves.

Past papers that receive many citations from papers published in a given year will form visible peaks in the forgetting curve of this year.
Plotting forgetting curves over several years, therefore, allows us to visually inspect how citations to milestone papers have evolved over time.
    To this end, the forgetting curve is read along a vertical line for a given year. If peaks form along this vertical line when read from top to bottom, milestone papers are being discovered and cited over time.
    On the other hand, if peaks vanish between the curves when read from top to bottom, it is an indication that milestone papers published in a given year are being ``forgotten'' in later years.
    Forgotten, here, refers to a decline in citations over time.
    Of course, the paper may still continue to receive citations (as most papers do over time). But in relative terms, it is being forgotten.
In this work, we apply forgetting curves to the ACM CHI proceedings.




\section{Method}%
\label{sec:method}%

\subsection{Data Collection}%
\label{sec:data-collection}%
The data for this analysis was 
collected from the ACM Digital Library (ACM--DL).
For each research article in the CHI proceedings from 1981--2024 (with exception of 1984), we downloaded the article's list of references, as listed on the article's page in the ACM--DL.
We exclude keynotes, panel sessions, and abstract only articles.
All information collected for this study is publicly accessible on the ACM--DL.
The resulting dataset consists of \NUMARTICLES~articles and a total of 629,120~references.
From this set of articles and references, we identify the milestone papers cited by the CHI community,
\transition{as described in the following section.}%


\subsection{Milestones}%
\label{sec:method:milestones-authors}%
\transition{The following sections describe how we identify milestones, their types and contributions, and the milestone authors.}%
%
\subsubsection{Milestone definition and identification}%
\label{sec:method:milestones}%
%
As mentioned in Section \ref{sec:rel:milestones}, there is no standard method for identifying milestone papers and different approaches have been used in prior work.
In our work, the papers with the highest total number of citations from CHI papers (over all CHI proceedings) are considered the milestones of the CHI community.
More specifically, we adopt the approach by \citeauthor{Reisz_2022_New_J._Phys._24_053041.pdf} \cite{Reisz_2022_New_J._Phys._24_053041.pdf} and 
define the top 30 most cited publications as milestones in our initial analysis.
As in \citeauthor{Reisz_2022_New_J._Phys._24_053041.pdf}, we acknowledge this choice is arbitrary.
However, to solidify our analysis we later extend our analysis to the top-100, top-300, and top-3000 papers cited in the CHI community, finding that the results are consistent for larger groups of milestones.

We further define ``super milestones'' as high performing milestones which have received more than 0.1\% (1 in 1000) of all citations in at least one proceedings year.
We derived this threshold empirically from visual inspection of citations over time (cf. \autoref{fig:milestonecitations}).
While this threshold is also arbitrary, it serves as a meaningful decision boundary for discussing different types of milestones in our analysis.

Note that in our analysis and throughout this whole article,
milestones are not limited to being published at CHI, but can also appear in other publication venues.
    A milestone can be a conference paper, journal article, or whole book.
    For simplicity, we refer to them as `milestones' or `milestone papers' in the remainder of this work.
Further note that the citation counts presented in our analysis (cf. \autoref{tab:milestones}) differ significantly from those that include all citations to a milestone, such as citation counts reported by Google Scholar.
Our focus in this work rests exclusively on the CHI community and how it cites its milestones (published at CHI or any other venues).

To identify the milestones, we parse each reference text to extract the publication year and the title with an iteratively developed set of over 100~regular expressions and logical checks.
We apply basic normalization to the paper's title, including lowercasing as well as removal of spaces, line breaks, and special characters.
By concatenating the year and normalized title into a unique ID, we can robustly identify papers by this ID (though, see Section \ref{sec:limitations} for limitations of this approach).
To answer RQ1, we can then analyze the citations from CHI papers to the uniquely identified papers.%
%
%
\subsubsection{Milestone types and contributions}%
\label{sec:method:milestone-types}%
To answer RQ2, we identify different types of milestones and analyze their contributions.
To identify types of milestones, we performed a cluster analysis, based on the data from the citation time series of the top-100 milestones most cited by CHI papers (over all proceedings years).
We triangulated this problem with different clustering approaches, including Ward’s method~\cite{WARD} and dynamic time warping with k-means~\cite{PETITJEAN2011678}.
The latter is a clustering technique that aligns and groups time series data with varying lengths 
by measuring similarity through optimal time alignment, allowing patterns to be identified despite temporal distortions.
This technique was used for clustering milestones in \autoref{tab:milestones} and \autoref{fig:milestonecitations}.
We explored the optimal number of clusters with the elbow criteria \cite{elbow} and other methods, as listed in~\cite{10.1145/3606274.3606278}.
However, analysis of dendrogram plots proved to provide the most reasonable and meaningful results, pointing to a three-cluster solution.

%

To identify the milestone contributions,
we inductively 
coded the key contributions of the top-30 milestones most cited in the CHI community.
To this end, the first author read the milestones' title and abstract and iteratively coded the milestones.
The codes were discussed in two meetings with the co-author to reach consensus.
This approach ensured consistency and alignment in the interpretation of the data.
One code was assigned to each milestone paper.
We acknowledge that a milestone may make more than one contribution.
However, our coding aimed to capture the key contribution of the community's top-performing milestones, answering the question of \textit{``what is the milestone most known and cited for?''}
The identified set of contributions was then mapped to existing contribution types by \citeauthor{2907069.pdf}~\cite{2907069.pdf} and \citeauthor{chi2024}~\cite{chi2024}.%
%
%
%
\subsubsection{Author identification}%
\label{sec:attachment-kernel}%
We parse author names from the ACM--DL's citation text and the scraped CHI paper references with a combination of regular expressions and logical checks to identify the author names in different reference notation styles (e.g., \textit{``Firstname Lastname. 2023.''} or \textit{``Lastname, Initials (2023a)``}).
Typically, these notation styles are not mixed within papers and individual references. This allows us to relatively reliably detect the pattern of author names and extract the author names from the CHI papers' references with regular expressions. 

For uniquely identifying authors, we follow \citeauthor{1520340.1520364.pdf}'s approach \cite{1520340.1520364.pdf} and base our analysis on only the author's last name and the first initial. While this may lead to author names being conflated, abbreviating first names is necessary because it constitutes the lowest common denominator in the different references formats found in CHI papers.
We discuss further limitations to this approach in Section~\ref{sec:limitations}.%
%
%
%
%
%
%
\subsection{Forgetting Curve Plots}%
\label{sec:method:forgettingcurves}%

We plot forgetting curves for the available CHI Proceedings years.
The construction of forgetting curves is straight-forward from the 
    collected 
data, using the formula given in Section \ref{sec:rel:forgettingcurves}.
The forgetting curves include citations to papers published at CHI and other venues. Therefore, the curves reach far back in time, but for the purpose of analyzing CHI, we limit the plot to our time frame of analysis (1981--2024).
On the secondary axis (in \autoref{fig:teaser}), we plot the total number of citations to the 30 most cited milestones in their respective publication year as vertical bars. The bars help in visually identifying some of the milestone peaks in the forgetting curves.%
%
%
\subsection{Milestone Coefficient}%
\label{sec:method:clustering}%

We define the milestone coefficient as a measure for the growing or diminishing impact of a paper relative to other papers published within a research community.
The aim of this metric is to classify papers on a continuous open-ended ``milestone spectrum.''
We design the metric to amplify differences between milestones that were popular once, but are being forgotten, and milestones that continue to grow in citations.
Our aim with this metric is to provide a tool for analyzing highly-cited scientific papers 
in a research community.
The metric goes beyond a
    discrete typology
of milestones and instead uses a continuous spectrum to measure the success of a paper in a research community.
To this end, we first define the trend of the citation series $c$ of a paper:
\[
\text{SIGN} = \begin{cases}
1, & \text{if } m \geq 0 \\
-1, & \text{if } m < 0
\end{cases}
\]
\[
m = 
\begin{cases}
\text{slope of the linear regression of } c \text{ on } x, & \text{if } n > 1 \\
0, & \text{if } n = 1
\end{cases}
\]
where
$c$ is the citation series
\[
c = \{c_0, c_1, \dots, c_{n-1}\}
\]
and each $c_i$ is the citation count for year $x_i$
and
$n$ is the number of years in the series.
A milestone that is fading from the collective memory of the research community will have a negative SIGN value, indicating a gradual decline in citations to this milestone. One the other hand, a milestone that is growing in citations has a positive SIGN value.
If there is only one citation value in the series ($n=1$), we define the slope $m=0$ and $SIGN=1$.
A paper's Total Citation Span (TCS) is a measure of the spread of citation counts over time. It is calculated as the difference between the highest and lowest citation counts observed for the paper: 
\[
\text{TCS} = 
\max(c) - \min(c)
\]
where $c$ is the series of citations to the paper over time.
TCS reflects the range in how frequently a paper has been cited, indicating the extent to which its citation count fluctuates. A higher TCS suggests greater variability in citation counts, while a lower TCS indicates a more consistent citation frequency.
%
We adjust the TCS by the paper's exponentially-scaled age span to account for the time that a paper had to accumulate citations over the years $x$ and derive the Total Citation Impact (TCI) of a paper as
\[
\text{TCI} = \frac{\textrm{TCS}}{(\max(x) - \min(x) + 1) \cdot e^{\alpha}}
\]
where $x$ are the years in the citation series $c$
and $\alpha$ is an exponential factor that adjusts for the fact that citation accrual follows a power law.
For estimating the parameter $\alpha$, we analyze all citations ever made in the CHI proceedings. For each cited paper, we construct the citation series (citations over the years) and fit a powerlaw distribution to these series.
This results in an estimate of $\alpha = 3.63$.
TCI is the first component of the Milestone Coefficient.

The second component is the Normalized Citation Impact (NCI) which provides a  measure of a paper's citation performance relative to all other citations to papers in a proceedings years $x$.
NCI is defined as the sum of this relative impact:
\begin{equation*}
\text{NCI}_j = \sum_{x} \frac{\text{citations}_{jx}}{\sum_{i} \text{citations}_{ix}}
\end{equation*}

We define the Milestone Coefficient (MC) of a paper as
\[
\text{MC} = \text{SIGN}~\cdot~\sqrt{\text{TCI}^2~\cdot~\text{NCI}^2}
\]

\begin{figure*}[!thb]%
\centering%
  \includegraphics[width=\linewidth]{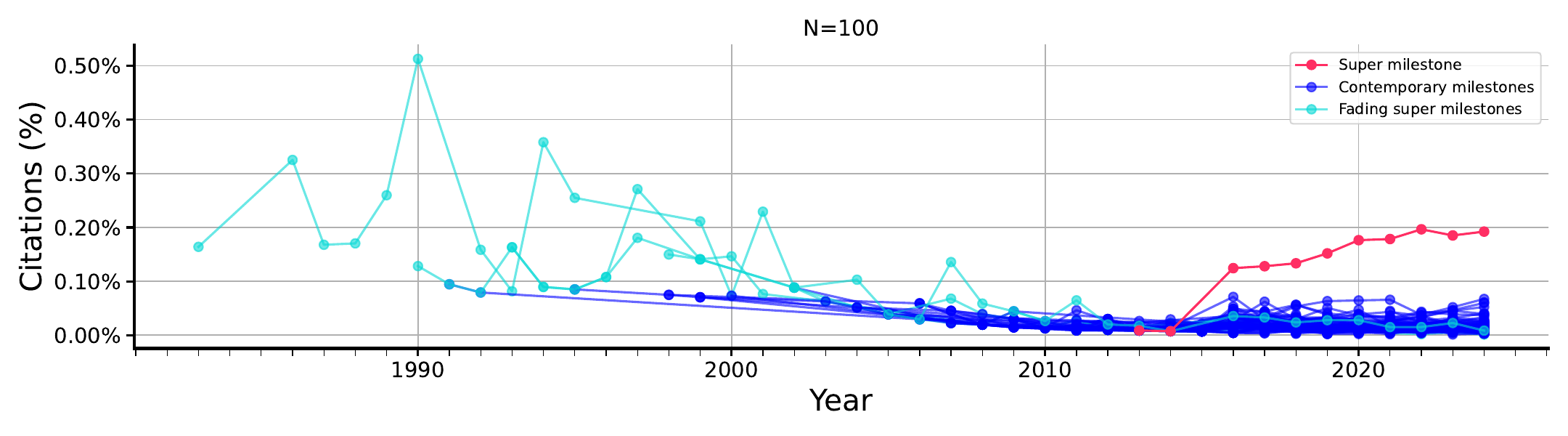}
  \caption{Relative number of citations to CHI's top-100 most cited milestones
  over time.
  A 
  clustering analysis
  reveals three types of milestones in CHI:
      one ``super milestone'' continues to grow in popularity (\textcolor{RED}{red}),
      some past super milestones are fading from the CHI community's memory (\textcolor{TURK}{light blue}),
      and some contemporary milestones are receiving continued attention, although at lower levels (\textcolor{BLUE}{blue}).}%
  \label{fig:milestonecitations}%
\end{figure*}%

The euclidean norm defines MC as the distance to the zero point (0,0) in a coordinate system 
spanned by TCI and NCI.
This can be explained as follows.
A high-performing milestone will show exceptional performance relative to other papers (NCI), but also 
when adjusted for its own age (TCI).
    The latter accounts for natural growth in citations that every paper experiences over its lifetime. This natural accrual of citations, however, does not necessarily make a paper a milestone. A high-performing milestone will demonstrate strong performance when compared to its own growth in citations (high TCI).
If either TCI or NCI is low, the overall MC is correspondingly low, emphasizing that both relative citation impact (NCI) and age-adjusted citation performance (TCI) are essential for a paper to be considered a milestone.
The euclidean norm reflects a paper's significance across both dimensions.
The milestone coefficient is deliberately not scaled to fall in the unit interval [-1, 1] to acknowledge the dynamic and evolving nature of citations.
Instead, a large positive milestone coefficient indicates a strongly performing and growing milestone, and a negative MC indicates a milestone in decline.
We apply the milestone coefficient to the 3,000~milestones most cited at CHI and plot the results later in Figure~\ref{fig:milestonecoefficient}.

\section{Results}%
\label{sec:results}%
\subsection{Milestones Cited by the CHI Community (RQ1 and RQ2)}%
\label{sec:milestones}%


The CHI community's 30 most cited milestones with year of publication and citation count are listed in \autoref{tab:milestones}.
Half of the 30 milestones most cited at CHI are published at the CHI Conference, the other half in other conferences, journals, or books.
The distribution of the CHI community's citations to milestones is long-tailed. Very few milestones received many citations, and there is a stark difference between CHI's top performing milestones and the rest of the milestones (cf. \autoref{tab:milestones}). This demonstrates how much the CHI community's attention rests on a few top-performing papers.
\autoref{tab:milestones} also highlights how relatively few citations the CHI community's milestones are receiving from CHI authors.
    For instance, the thirtieth paper in the top-30 list has received only 75~citations from the CHI community since its publication in 2006.
    Expanding this list to the top-300 most cited papers, the papers in 300th place each received only 34~citations from the CHI community (see the Supplemental Material).
\transition{The following section provides a description of CHI's top-30 most cited milestones.}

\subsubsection{Milestone description (RQ1)}

The 30~milestones listed in \autoref{tab:milestones} represent key contributions that have significantly influenced the landscape of HCI research.
%
%
Together, these milestones reflect the diverse
advances that have shaped HCI research and practice.
\transition{%
The following section presents a typology of milestones based on their citation performance.%
}%

\begin{table*}[tb]%
\caption{%
Milestones most cited in the ACM CHI Proceedings (1981--2024).
}%
\label{tab:milestones}%
\small%
\begin{tabularx}{\textwidth}{
    c
    c
    c
    >{\raggedright\arraybackslash}p{\dimexpr0.175\textwidth}
    X
    c
    l
    c
}%
    \toprule
    No. &
    Year &
    Type & 
    Milestone & Title & 
    Citations\textsuperscript{$\dagger$}
    & Source &
    MC$\times$10\textsuperscript{-2}
    \\
\midrule
    1. &
    2006 &
        \textcolor{RED}{\rule{0.5em}{0.5em}} &
        \citeauthor{M1} \cite{M1} &
        Using thematic analysis in psychology
         & 770 & journal
         & 7.68
         \\
    2. &
    1988 &
        \textcolor{BLUE}{\rule{0.5em}{0.5em}} &
        \citeauthor{M2} \cite{M2} &
        Development of {NASA-TLX} (task load index): results of empirical and theoretical research
        & 262 & book 
         & 2.08
        \\
    3. &
    2011 & 
    \textcolor{BLUE}{\rule{0.5em}{0.5em}} &
    \citeauthor{M3} \cite{M3} & The aligned rank transform for nonparametric factorial analyses using only ANOVA procedures
        & 179 & CHI 
         & 1.66
        \\
    4. &
    2007 & 
    \textcolor{BLUE}{\rule{0.5em}{0.5em}} &
    \citeauthor{M4} \cite{M4} &
        Research through design as a method for interaction design research in HCI 
        & 162 & book 
         & 1.38
        \\
    5. &
    2010 &
        \textcolor{BLUE}{\rule{0.5em}{0.5em}} &
        \citeauthor{M5} \cite{M5} &
        Feminist {HCI}: taking stock and outlining an agenda for design
        & 141 & CHI 
         & 1.19
        \\
    6. &
    2012 & 
        \textcolor{BLUE}{\rule{0.5em}{0.5em}} &
        \citeauthor{BRAUN} \cite{BRAUN} & 
        Thematic analysis
        & 129 & book
         & 1.85
        \\
    7. &
    2019 &
        \textcolor{BLUE}{\rule{0.5em}{0.5em}} &
        \citeauthor{MX} \cite{MX} &
        Reflecting on reflexive thematic analysis
         & 125 & book
         & 2.22
         \\
    8. &
    2006 &
        \textcolor{BLUE}{\rule{0.5em}{0.5em}} &
        \citeauthor{MX2} \cite{MX2} &
        NASA-task load index (NASA-TLX); 20 years later
         & 113 & journal
         & 1.35
         \\
    9. &
    2019 &
        \textcolor{BLUE}{\rule{0.5em}{0.5em}} &
        \citeauthor{M6} \cite{M6} &
        Reliability and inter-rater reliability in qualitative research: norms and guidelines for CSCW and HCI practice
         & 113 & CHI
         & 1.68
         \\
    10. &
    2010 &
        \textcolor{BLUE}{\rule{0.5em}{0.5em}} &
        \citeauthor{STAGE} \cite{STAGE} &
        A stage-based model of personal informatics systems
         & 107 & CHI
         & 1.01
         \\
    11. &
    2003 & 
        \textcolor{BLUE}{\rule{0.5em}{0.5em}} &
        \citeauthor{TECHPROBE} \cite{TECHPROBE} &
        Technology probes: inspiring design for and with families
        & 106 & CHI 
         & 0.98
        \\
    12. &
    2005 &
        \textcolor{BLUE}{\rule{0.5em}{0.5em}} &
        \citeauthor{REVD} \cite{REVD}
        & Reflective design
        & 97 & CC 
         & 0.82
        \\
    13. &
    1996 &
        \textcolor{BLUE}{\rule{0.5em}{0.5em}} &
        \citeauthor{SUS} \cite{SUS} &
        SUS: a `quick and dirty' usability scale
        & 94 & book
         & 0.98
        \\
    14. &
    2003 & 
        \textcolor{BLUE}{\rule{0.5em}{0.5em}} &
        \citeauthor{AMB} \cite{AMB} &
        Ambiguity as a resource for design
        & 92 & CHI
         & 0.76
        \\
    15. &
    2017 &
        \textcolor{BLUE}{\rule{0.5em}{0.5em}} &
        \citeauthor{INTER} \cite{INTER} &
        Intersectional HCI: engaging identity through gender, race, and class 
         & 86 & CHI
         & 0.90
         \\
    16. &
    1997 & 
        \textcolor{TURK}{\rule{0.5em}{0.5em}} &
        \citeauthor{BITS} \cite{BITS} &
        Tangible bits: towards seamless interfaces between people, bits and atoms
        & 84 & CHI
         & 0.77
        \\
    17. &
    2010 &
        \textcolor{BLUE}{\rule{0.5em}{0.5em}} &
        \citeauthor{SUSHCI} \cite{SUSHCI} &
        Mapping the landscape of sustainable HCI
        & 84 & CHI
         & 0.85
        \\
    18. &
    2010 &
        \textcolor{BLUE}{\rule{0.5em}{0.5em}} &
        \citeauthor{POSTCO} \cite{POSTCO} &
        Postcolonial computing: a lens on design and development
        & 84 & CHI
         & 0.77
        \\
    19. &
    2014 &
        \textcolor{BLUE}{\rule{0.5em}{0.5em}} &
        \citeauthor{PERCO} \cite{PERCO} &
        Personal tracking as lived informatics
        & 84 & CHI
         & 0.59
        \\
    20. &
    2019 &
        \textcolor{BLUE}{\rule{0.5em}{0.5em}} &
        \citeauthor{AI} \cite{AI} &
        Guidelines for human-AI interaction
        & 84 & CHI
         & 0.60
        \\
    21. &
    2013 &
        \textcolor{BLUE}{\rule{0.5em}{0.5em}} &
        \citeauthor{DREAM} \cite{DREAM} &
        Speculative everything: design, fiction, and social dreaming
        & 83 & book
         & 0.96
        \\
    22. &
    1999 &
        \textcolor{BLUE}{\rule{0.5em}{0.5em}} &
        \citeauthor{DCU} \cite{DCU} &
        Design: Cultural probes
        & 80 & journal 
         & 0.73
        \\
    23. &
    2015 &
        \textcolor{BLUE}{\rule{0.5em}{0.5em}} &
        \citeauthor{UBI} \cite{UBI} &
        A lived informatics model of personal informatics
        & 78 & UbiComp
         & 0.78
        \\
    24. &
    2016 &
        \textcolor{BLUE}{\rule{0.5em}{0.5em}} &
        \citeauthor{HAPTIC} \cite{HAPTIC} &
        Haptic retargeting: dynamic repurposing of passive haptics for enhanced virtual reality experiences
        & 78 & CHI 
         & 0.65
        \\
    25. &
    2009 &
        \textcolor{BLUE}{\rule{0.5em}{0.5em}} &
        \citeauthor{GEST} \cite{GEST} &
        User-defined gestures for surface computing
        & 77 & CHI 
         & 0.65
        \\
    26. &
    2013 &
        \textcolor{BLUE}{\rule{0.5em}{0.5em}} &
        \citeauthor{INFORM} \cite{INFORM} &
        inFORM: dynamic physical affordances and constraints through shape and object actuation
        & 77 & UIST
         & 0.75
        \\
    27. &
    1993 &
        \textcolor{BLUE}{\rule{0.5em}{0.5em}} &
        \citeauthor{SIMSICK} \cite{SIMSICK} &
        Simulator sickness questionnaire: an enhanced method for quantifying simulator sickness
        & 76 & journal
         & 0.62
        \\
    28. &
    2011 &
        \textcolor{BLUE}{\rule{0.5em}{0.5em}} &
        \citeauthor{D3} \cite{D3} &
        D3: Data-driven documents
        & 76 & journal
         & -0.62
        \\
    29. &
    2003 &
        \textcolor{BLUE}{\rule{0.5em}{0.5em}} &
        \citeauthor{MXX} \cite{MXX} &
        Phrase sets for evaluating text entry techniques
        & 75 & CHI
         & 0.57
        \\
    30. &
    2006 &
        \textcolor{BLUE}{\rule{0.5em}{0.5em}} &
        \citeauthor{GROUND} \cite{GROUND} &
        Constructing grounded theory a practical guide through qualitative analysis
        & 75 & book
         & 0.77
        \\
\bottomrule%
\end{tabularx}%
\\
\begin{minipage}{\textwidth}
\raggedright
\scriptsize%
~ \textsuperscript{$\dagger$}
Since our focus is on the CHI community, the citation count listed in this table only includes  citations originating from CHI papers.
\end{minipage}
\end{table*}%

\begin{table*}[thb]%
\caption{Contributions of the 30 milestones most-cited by the CHI community.}%
\label{tab:milestonedescription}%
\small%
\begin{tabularx}{\textwidth}{
    c
    c
    >{\raggedright\arraybackslash}p{\dimexpr0.17\textwidth}
    X
    l
}
\toprule
    No. &
    Year & Milestone & Description & Contribution \\
\midrule
    1. &
    \citeyear{M1} &
    \textcolor{RED}{\rule{0.5em}{0.5em}}
    \citeauthor{M1} \cite{M1}  &
    \footnotesize
    Thematic Analysis, 
    a method for analyzing qualitative data
    & method
    \\

    2. &
    \citeyear{M2} &
    \textcolor{BLUE}{\rule{0.5em}{0.5em}}
    \citeauthor{M2} \cite{M2}  &
    \footnotesize
    NASA Task Load Index (NASA-TLX), a 
    tool for measuring perceived workload
    & method/tool
    \\

    3. &
    \citeyear{M3} &
    \textcolor{BLUE}{\rule{0.5em}{0.5em}}
    \citeauthor{M3} \cite{M3}  & 
    \footnotesize
    Aligned Rank Transform (ART) 
    for non-parametric factorial analyses through ANOVA procedures
    & method
    \\

    4. &
    \citeyear{M4} &
    \textcolor{BLUE}{\rule{0.5em}{0.5em}}
    \citeauthor{M4} \cite{M4}  & 
    \footnotesize 
    Research through Design as a way to generate knowledge by reflecting on design practice
    & method
    \\

    5. &
    \citeyear{M5} &
    \textcolor{BLUE}{\rule{0.5em}{0.5em}}
    \citeauthor{M5} \cite{M5}  & 
    \footnotesize
    Feminist HCI, 
    a critical agenda that considers gender, power, and identity in design
    & framework/agenda
    \\

    6. &
    \citeyear{BRAUN} &
    \textcolor{BLUE}{\rule{0.5em}{0.5em}}
    \citeauthor{BRAUN} \cite{BRAUN}  & 
    \footnotesize
    Thematic Analysis,
    a method for analyzing qualitative data
    & method/guidelines
    \\

    7. &
    \citeyear{MX} &
    \textcolor{BLUE}{\rule{0.5em}{0.5em}}
    \citeauthor{MX} \cite{MX}  & 
    \footnotesize
    Reflexive Thematic Analysis, a clarification of \citeauthor{M1}'s initial method
    & method/guidelines
    \\

    8. &
    \citeyear{MX2} &
    \textcolor{BLUE}{\rule{0.5em}{0.5em}}
    \citeauthor{MX2} \cite{MX2}  & 
    \footnotesize
    A reflection on the NASA-TLX's two decades of impact
    & method/tool
    \\

    9. &
    \citeyear{M6} &
    \textcolor{BLUE}{\rule{0.5em}{0.5em}}
    \citeauthor{M6} \cite{M6}  & 
    \footnotesize
    Guidelines on reliability and inter-rater reliability in qualitative research
    & guidelines
    \\

    10. &
    \citeyear{STAGE} &
    \textcolor{BLUE}{\rule{0.5em}{0.5em}}
    \citeauthor{STAGE} \cite{STAGE}  & 
    \footnotesize
    A stage-based model of personal informatics systems
    & model
    \\

    11. &
    \citeyear{TECHPROBE} &
    \textcolor{BLUE}{\rule{0.5em}{0.5em}}
    \citeauthor{TECHPROBE} \cite{TECHPROBE}  & 
    \footnotesize
    Technology Probes, an expansion of Cultural Probes
    & method
    \\

    12. &
    \citeyear{REVD} &
    \textcolor{BLUE}{\rule{0.5em}{0.5em}}
    \citeauthor{REVD} \cite{REVD}  & 
    \footnotesize
    Reflective Design, 
    an approach to question underlying assumptions in technology
    & framework
    \\

    13. &
    \citeyear{SUS} &
    \textcolor{BLUE}{\rule{0.5em}{0.5em}}
    \citeauthor{SUS} \cite{SUS}  &
    \footnotesize
    System Usability Scale (SUS), 
    a 
    tool for assessing usability
    & method/tool
    \\

    14. &
    \citeyear{AMB} &
    \textcolor{BLUE}{\rule{0.5em}{0.5em}}
    \citeauthor{AMB} \cite{AMB}  &
    \footnotesize
    Ambiguity as a Resource for Design
    & framework/concept
    \\

    15. &
    \citeyear{INTER} &
    \textcolor{BLUE}{\rule{0.5em}{0.5em}}
    \citeauthor{INTER} \cite{INTER}  & 
    \footnotesize 
    Intersectional HCI, a critical perspective advocating for designs that address the complexities of gender, race, and class
    & framework
    \\

    16. &
    \citeyear{BITS} &
    \textcolor{TURK}{\rule{0.5em}{0.5em}}
    \citeauthor{BITS} \cite{BITS}  & 
    \footnotesize
    Tangible Bits, a 
    vision for seamless interfaces that blend the digital and physical worlds, laying the groundwork for tangible interaction
    & concept
    \\

    17. &
    \citeyear{SUSHCI} &
    \textcolor{BLUE}{\rule{0.5em}{0.5em}}
    \citeauthor{SUSHCI} \cite{SUSHCI}  &
    \footnotesize
    Sustainable HCI, urging the field to consider environmental impacts
    & framework
    \\

    18. &
    \citeyear{POSTCO} &
    \textcolor{BLUE}{\rule{0.5em}{0.5em}}
    \citeauthor{POSTCO} \cite{POSTCO}  &
    \footnotesize
    Postcolonial Computing, a framework for examining design and development through postcolonial theory
    & framework
    \\

    19. &
    \citeyear{PERCO} &
    \textcolor{BLUE}{\rule{0.5em}{0.5em}}
    \citeauthor{PERCO} \cite{PERCO}  & 
    \footnotesize
    The concept of ``lived informatics'', examining how people integrate personal tracking into their daily lives
    & model
    \\

    20. &
    \citeyear{AI} &
    \textcolor{BLUE}{\rule{0.5em}{0.5em}}
    \citeauthor{AI} \cite{AI}  & 
    \footnotesize
    Guidelines for Human-AI Interaction, offering best practices for designing user-centric AI systems
    & guidelines
    \\

    21. &
    \citeyear{DREAM} &
    \textcolor{BLUE}{\rule{0.5em}{0.5em}}
    \citeauthor{DREAM} \cite{DREAM}  & 
    \footnotesize
    Speculative Design, a method using fictional scenarios to explore alternative futures
    & framework/concept
    \\

    22. &
    \citeyear{DCU} &
    \textcolor{BLUE}{\rule{0.5em}{0.5em}}
    \citeauthor{DCU} \cite{DCU}  & 
    \footnotesize
    Cultural Probes, a method using evocative artifacts to engage people in the design process
    & method
    \\

    23. &
    \citeyear{UBI} &
    \textcolor{BLUE}{\rule{0.5em}{0.5em}}
    \citeauthor{UBI} \cite{UBI}  & 
    \footnotesize
    Lived Informatics Model, a development of \citeauthor{PERCO}'s method
    & model
    \\

    24. &
    \citeyear{HAPTIC} &
    \textcolor{BLUE}{\rule{0.5em}{0.5em}}
    \citeauthor{HAPTIC} \cite{HAPTIC}  &
    \footnotesize
    Haptic Retargeting, a technique for creating more immersive interactions in VR experiences
    & method
    \\        

    25. &
    \citeyear{GEST} &
    \textcolor{BLUE}{\rule{0.5em}{0.5em}}
    \citeauthor{GEST} \cite{GEST}  & 
    \footnotesize
    User-defined gestures, underscoring the importance of user preferences in surface computing interactions
    & method
    \\

    26. &
    \citeyear{INFORM} &
    \textcolor{BLUE}{\rule{0.5em}{0.5em}}
    \citeauthor{INFORM} \cite{INFORM}  &
    \footnotesize 
    inFORM, a shape display that allows for dynamic physical affordances through actuation, demonstrating the potential of tangible interfaces in HCI
    & method
    \\

    27. &
    \citeyear{SIMSICK} &
    \textcolor{BLUE}{\rule{0.5em}{0.5em}}
    \citeauthor{SIMSICK} \cite{SIMSICK}  &
    \footnotesize
    Simulator Sickness Questionnaire, a method for quantifying simulator sickness
    & method/tool
    \\

    28. &
    \citeyear{D3} &
    \textcolor{BLUE}{\rule{0.5em}{0.5em}}
    \citeauthor{D3} \cite{D3}  & 
    \footnotesize
    D3 (Data-Driven Documents), a 
    software library for creating interactive web visualizations
    & tool
    \\    

    29. &
    \citeyear{MXX} &
    \textcolor{BLUE}{\rule{0.5em}{0.5em}}
    \citeauthor{MXX} \cite{MXX}  &
    \footnotesize
    Standardized phrase sets for evaluating text entry techniques 
    as benchmark for assessing typing performance
    & method/tool
    \\
    
    30. &
    \citeyear{GROUND} &
    \textcolor{BLUE}{\rule{0.5em}{0.5em}}
    \citeauthor{GROUND} \cite{GROUND}  & 
    \footnotesize
    A practical guide to Constructing Grounded Theory
    & method
    \\
                                
\bottomrule
\end{tabularx}%
\end{table*}%

\label{sec:milestones-types}%
\autoref{fig:milestonecitations} tracks citations from the CHI community to its 100 most cited milestones over time,
relative to all citations made in a given CHI proceedings year.
Our cluster analysis of these 100 citation series finds three different types of milestones being cited at CHI:%
%
\begin{enumerate}%
    \item \textcolor{RED}{\rule{0.5em}{0.5em}} \textit{Super milestones} (1\%):
    One milestone continues to grow in popularity, although its citations from CHI authors may be approaching saturation in recent years.
    In the 2024 CHI proceedings, this milestone received 0.2\% of all citations made in CHI papers.
    \textit{Example:}
    \begin{itemize}%
        \item[\textendash]
            \citeauthor{M1}'s Thematic Analysis method \cite{M1}
    \end{itemize}%
    
    \item \textcolor{TURK}{\rule{0.5em}{0.5em}} \textit{Fading super milestones} (4\%):
    Four milestones were very popular in the past, with one receiving over 0.5\% of all citations in one proceedings year. However, these milestone have been fading from the CHI community's memory.
    \textit{Examples:}
     \begin{itemize}%
         \item[\textendash]
            \citeauthor{10.5555/578027}'s  
            ``The  Psychology of Human-Computer Interaction'' \cite{10.5555/578027}
         \item[\textendash]
            \citeauthor{BITS}'s ``Tangible Bits'' \cite{BITS}
         \item[\textendash]
            \citeauthor{FITTSDESIGNTOOL}'s ``Fitts' law as a research and design tool'' \cite{FITTSDESIGNTOOL}
     \end{itemize}%
    
    \item \textcolor{BLUE}{\rule{0.5em}{0.5em}} \textit{%
    Milestones} (95\%):
    Some milestones are receiving continued attention, although at levels much lower than prior super milestones.
    \textit{Examples:}
    \begin{itemize}%
        \item[\textendash]
            \citeauthor{M4}'s Research Through Design 
            \cite{M4}
        \item[\textendash]
            \citeauthor{M2}'s NASA TLX \cite{M2}
        \item[\textendash]
            \citeauthor{M5}'s Feminist HCI \cite{M5}
    \end{itemize}%
\end{enumerate}


\transition{%
Next, we turn our attention to analyzing the different contribution types made by the milestones.
This is crucial for understanding what type of work
is being appreciated in the long run.}

\subsubsection{Milestone contributions (RQ2)}%
\label{sec:milestone-contributions}%



\autoref{tab:milestonedescription} provides a brief description of the unique contribution of each of the 30 milestones most cited at CHI.
Our analysis identifies seven key types of milestone contributions:
    method, tool, framework, agenda, guidelines, model, and concept (see \autoref{tab:milestones} and \autoref{tab:contribution-comparison}).
These seven categories encapsulate the key ways in which the top-30 milestones have impacted and advanced the field of HCI.

We compare the key contribution types of the 30 milestones with \citeauthor{2907069.pdf}'s seven types of research contributions in HCI \cite{2907069.pdf} and the Contribution Types for CHI 2024 \cite{chi2024} (see \autoref{tab:contribution-comparison}).
The most prominent contribution shared with existing typologies is the method contribution, which directly corresponds to \citeauthor{2907069.pdf}'s methodological contribution.
    More than half of the top-30 milestones make methodological contributions (see \autoref{tab:milestonedescription}).

However, many other contribution types in HCI are not represented in the CHI community's milestone contributions.
For instance, prototypes and systems are a common type of contribution in HCI research, with several hundred papers designing, implementing, and evaluating systems and prototypes each year.
However, this type of contribution is not represented in the list of key milestone contributions.
Also not represented are less common contribution types, such as 
literature survey, dataset, and replication.%
%
%
\begin{table*}[!htb]%
\caption{Comparison of milestone contribution types with existing contribution types in HCI.
}%
\label{tab:contribution-comparison}%
\begin{tabularx}{\linewidth}{
    l
    X
    >{\raggedright\arraybackslash}p{1.8cm}
    >{\raggedright\arraybackslash}p{2cm}
}%
    \toprule
    \small Contribution
    & \small Description with examples
        & \small \citeauthor{2907069.pdf} (\citeyear{2907069.pdf})\textsuperscript{$*$}
        & \small \citeauthor{chi2024}
        (\citeyear{chi2024})\textsuperscript{$\dagger$}
    \\
\midrule
    \small Method
    & \small A systematic approach or technique used for conducting research or analysis in HCI.
        \textit{%
        Examples:
        qualitative methods (e.g., thematic analysis), quantitative methods (e.g., aligned rank transform)
        }%
    & \small Methodological
    & \small 
    Methodology
        \\
        
    \small Tool 
    & \small A practical instrument or software designed to perform specific tasks or measurements in HCI studies. 
        \textit{%
        Examples:
        survey tools (e.g., NASA-TLX, SUS),
        software libraries (e.g., D3)
        }%
    & \small Artifact
    & \small 
            Artifacts
        \\
        
    \small Framework
    & \small A comprehensive structure outlining key concepts in HCI which organizes knowledge and provides a structured approach for HCI research. 
        \textit{%
        Examples:
            Feminist HCI, Reflective Design
        }%
    & \small Theoretical
    & \small 
    Theory
        \\
        
    \small Agenda
    & \small A call for future research or practice directions in HCI.
        \textit{%
        Example:
            Feminist HCI agenda
        }%
    & \small Opinion
    & \small 
    Argument 
        \\
        
    \small Guidelines
    & \small Prescriptive recommendations or best practices for designing or evaluating human-computer interaction. 
        \textit{%
        Example:
        Guidelines for inter-rater reliability
        }%
    & \small Methodological
    & \small 
    Methodology
        \\
        
    \small Model 
    & \small An abstract representation detailing the components and processes of a system or phenomenon in HCI. 
        \textit{%
        Example:
        Model of personal informatics
        }%
    & \small Theoretical
    & \small 
    Theory
        \\
        
    \small Concept 
    & \small A foundational idea or theory that introduces new perspectives or approaches to HCI design and research.
        \textit{%
        Example:
        Ambiguity
        }%
    & \small Theoretical
    & \small 
    Theory
        \\
        
\bottomrule%
\end{tabularx}%
\vspace{-.5\baselineskip}%
\\
\begin{tabularx}{\linewidth}{X}%
    \raggedright
    \footnotesize
    ~ \textsuperscript{$*$} Research contribution types in HCI \cite{2907069.pdf}: Empirical, Artifact, Methodological, Theoretical, Dataset, Survey, Opinion
\\
    ~ \textsuperscript{$\dagger$} Contributions to CHI \cite{chi2024}: 
    1) Development or Refinement of Interface Artifacts or Techniques,
    2) Understanding Users,
    3) Systems, Tools, Architectures, and Infrastructure,
    4) Methodology,
    5) Theory,
    6) Innovation, Creativity, and Vision,
    7) Argument,
    8) Validation and Replication
\end{tabularx}%
\end{table*}%
%
%
%
\subsubsection{Case study: The rise of CHI's super milestone}%
\label{sec:supermilestone}%
\citeauthor{M1}’s Thematic Analysis \cite{M1} 
is a unique ``super'' milestone (cf. \autoref{fig:citationcurves} and \autoref{fig:milestonecoefficient}).
What makes this milestone stand out is not only its absolute citation count, but also its speed of adoption in the CHI community.
    \citeauthor{M1}'s method for qualitative analysis 
    was first published in the year \citeyear{M1} and was immediately useful to researchers in HCI, resulting in rapid adoption in the CHI community (cf. \autoref{fig:milestonecitations} and \cite{3544548.3581203.pdf}).
In fact, Thematic Analysis occupies not one but three of CHI's 30 most cited milestones (see Table \ref{tab:milestones}).
Together, these milestone papers have amassed
1024 
citations from CHI papers.
While this number 
pales in comparison to the total number of citations that these milestone papers have received from other sources, it does demonstrate the popularity of Thematic Analysis in the CHI community.
\citeauthor{M1}’s milestones have received almost four times more citations 
than the second most cited milestone at CHI (see \autoref{tab:milestones}).

In the year 2024, \citeauthor{M1}'s most-cited milestone \cite{M1} accounts for about 1 in 500 citations (0.2\%) in the CHI proceedings.
Given that CHI papers include an average of about 88~references in 2024~\cite{oppenlaender-citations-HCI}, this means we can estimate that, on average, about one in six (5.7) CHI papers cite \citeauthor{M1}'s work.
At present, no other milestone demonstrates such an exceptional performance.
\transition{In the following section, we examine how the CHI community's attention to this and other milestones has evolved over time.%
}%
%
%
%
%
%
%
%
\subsection{How has the CHI Community’s Attention to its Milestones evolved over time? (RQ3)}%
\label{sec:citations-to-milestones}%
Forgetting curves graphically depict the attention of the CHI community to past papers and milestones (see \autoref{fig:teaser}).
A visual inspection of the forgetting curves yields the insight that the forgetting curves have become ``smoother'' over time.
    Each year, there are fewer visible peaks in the forgetting curves and, thus, fewer milestones are being cited in the ACM CHI proceedings.
    In other terms, papers published today are less likely to cite older papers.
\begin{figure}[!htb]%
\centering%
  \includegraphics[width=\linewidth]{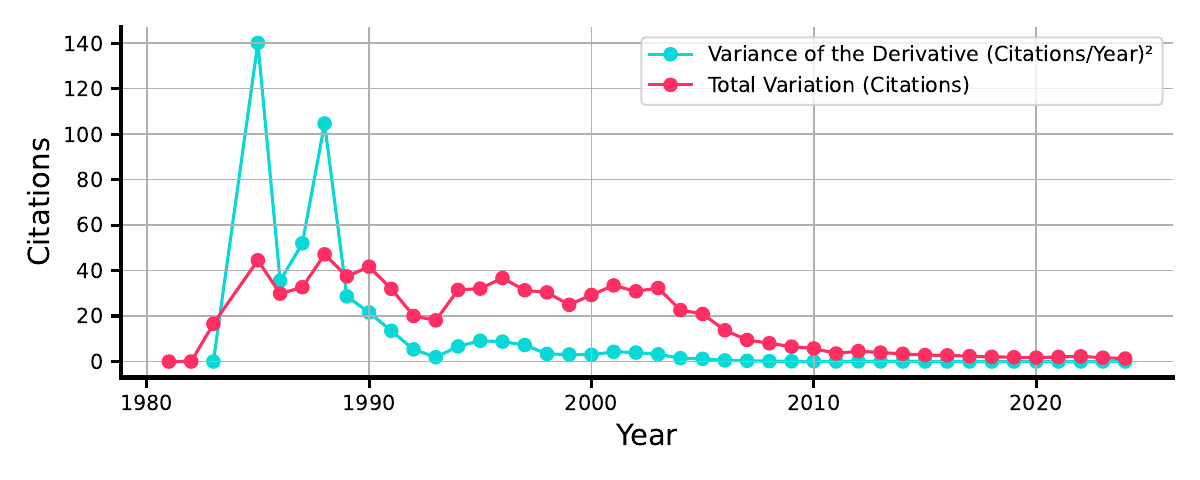}
  \caption{CHI's forgetting curves have become ``smoother'' over time, lacking the visible peaks that are characteristic of milestones.}
  \label{fig:smoothness}
\end{figure}%
We can quantify this visual effect by calculating and plotting the variance of the derivative and the total variation of each of the forgetting curves (see \autoref{fig:smoothness}).
    The variance of the derivative measures the
    variability of the rate of change (derivative) in the forgetting curves.
    The total variation represents the cumulative amount of change over the forgetting curve.
The plot in \autoref{fig:smoothness}
shows that the forgetting curves have become ``smoother'' over time, providing 
graphical evidence of  a fading memory of milestones in the CHI community.%

\begin{figure}[!htb]%
\centering%
  \includegraphics[width=\linewidth]{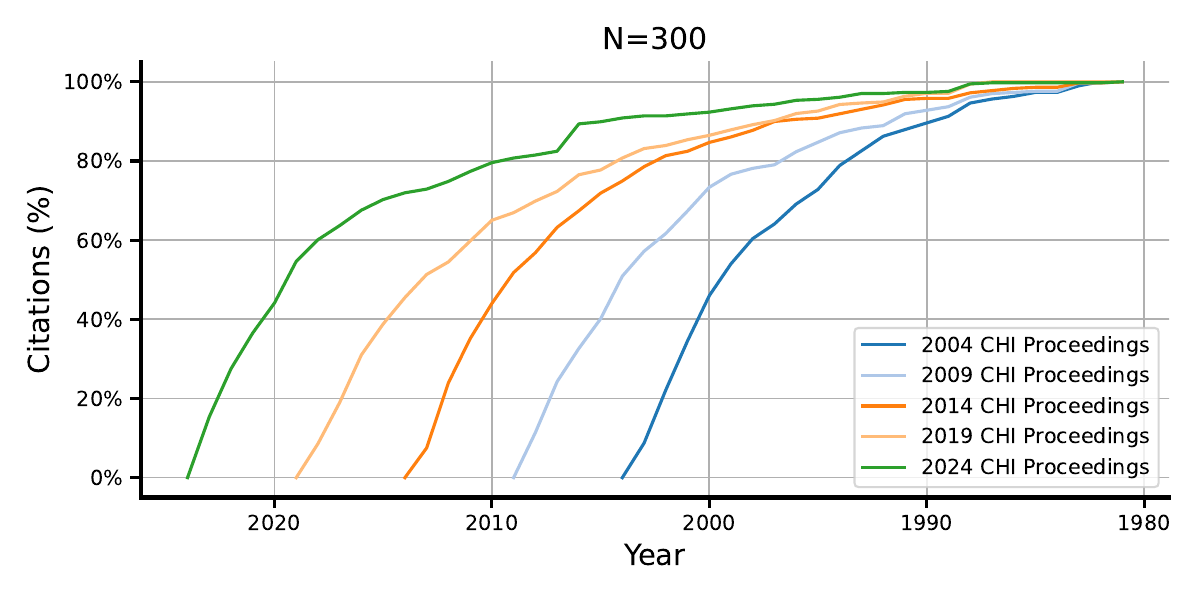}%
  \caption{Cumulative citations to the prior 300 most cited milestones from selected CHI proceedings years.}%
  \label{fig:cumulative-citations}
\end{figure}%
In the 2024 CHI Proceedings, a significant portion of the community's attention is directed toward more recent contributions.
To understand this trend, we calculated the cumulative percentage of citations from selected CHI proceedings to the 300 most cited milestones published in years prior (see \autoref{fig:cumulative-citations}).
The results highlight the concentration of citations to recent milestones.
    Notably, 15.4\% of the citations point to works published in 2023.
    This cumulative percentage increases to 27.5\% when including works from 2022 and 2023, and further rises to 60.1\% when considering all works from 2018 onwards.
    By 2010, the cumulative percentage reaches 79.6\%, and by 2005, it climbs to 89.9\%.
The bulk of the CHI community's attention rests on papers published in a time window of between 5 and 15 years.
This indicates a strong emphasis on citing more recent milestones within the CHI community.


\begin{figure*}[!htb]%
\centering%
  \includegraphics[width=.245\linewidth]{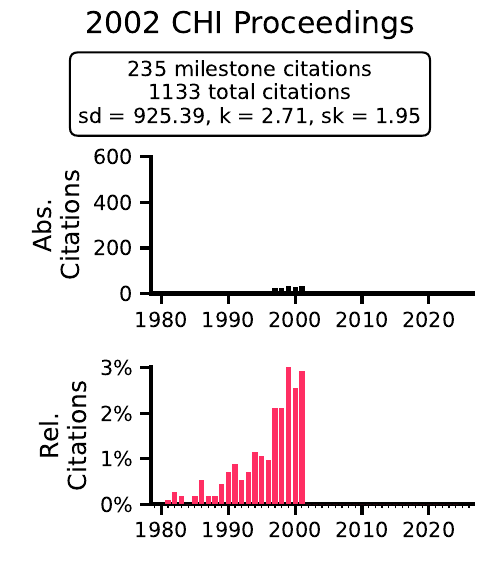}
  \includegraphics[width=.245\linewidth]{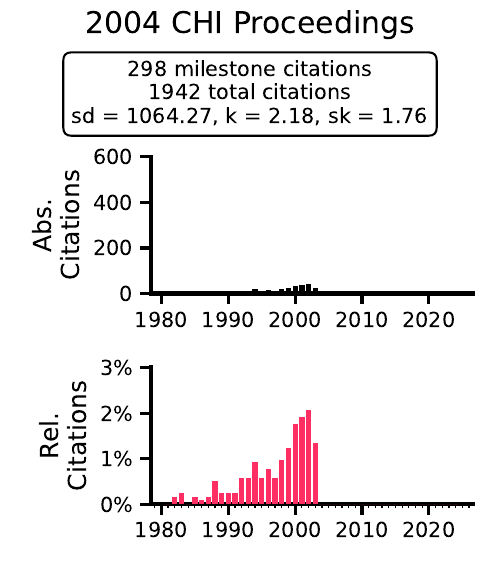}
  \includegraphics[width=.245\linewidth]{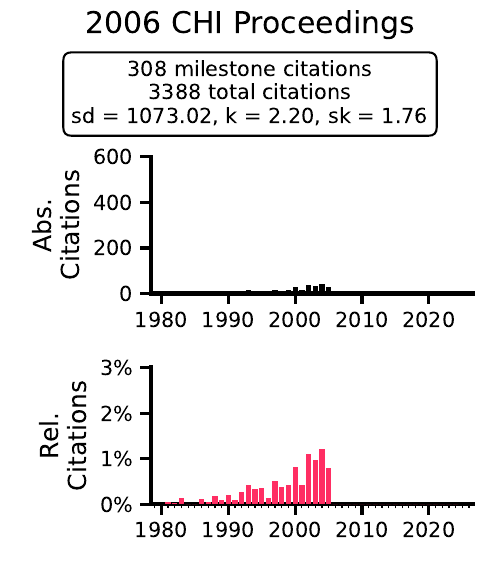}
  \includegraphics[width=.245\linewidth]{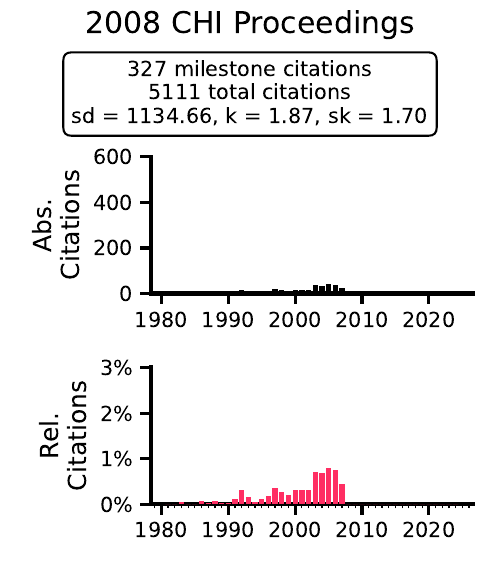}
  \\
  \includegraphics[width=.245\linewidth]{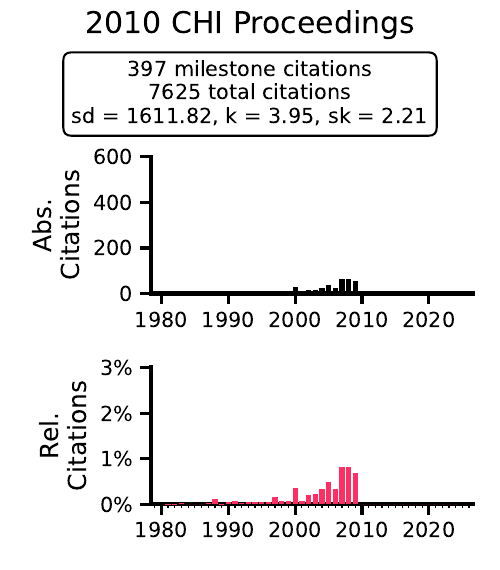}
  \includegraphics[width=.245\linewidth]{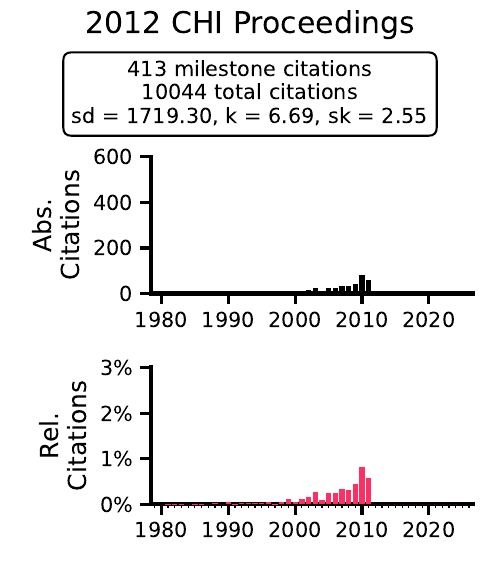}
  \includegraphics[width=.245\linewidth]{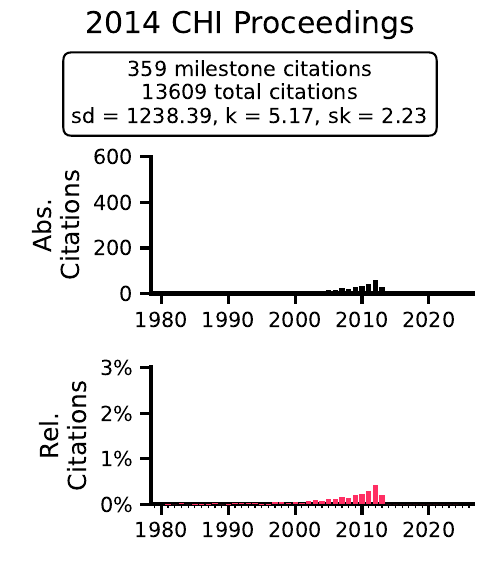}
  \includegraphics[width=.245\linewidth]{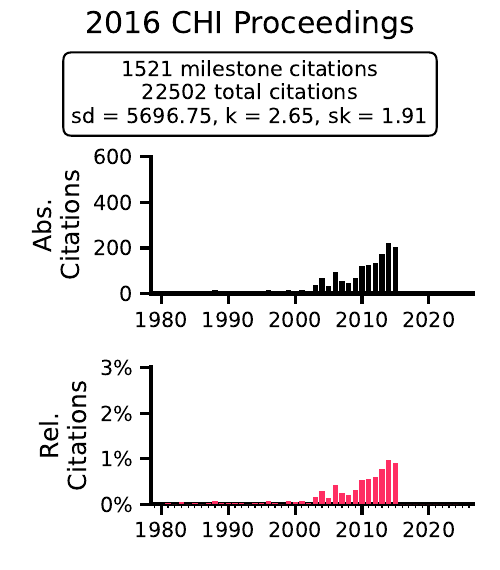}
  \\
  \includegraphics[width=.245\linewidth]{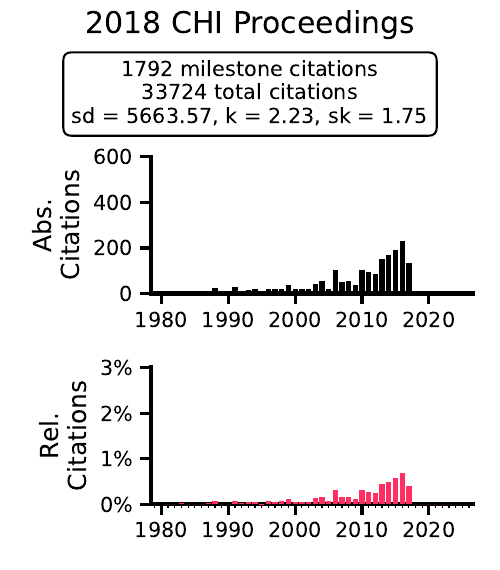}
  \includegraphics[width=.245\linewidth]{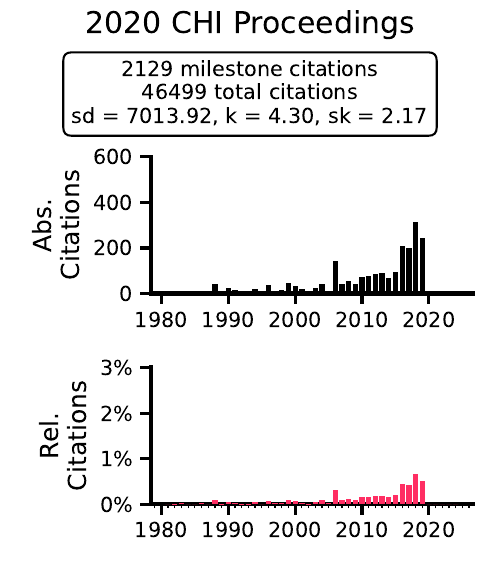}
  \includegraphics[width=.245\linewidth]{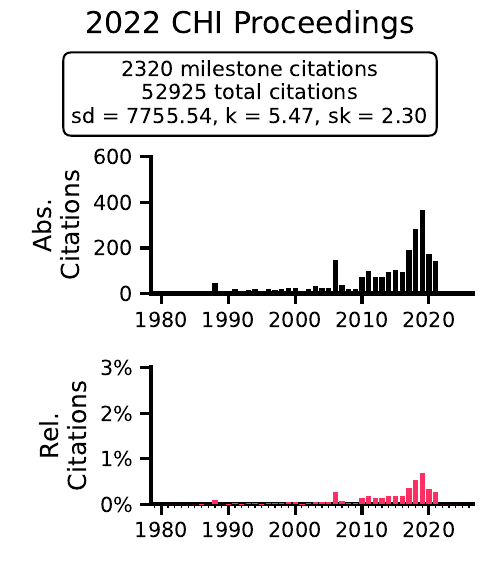}
  \includegraphics[width=.245\linewidth]{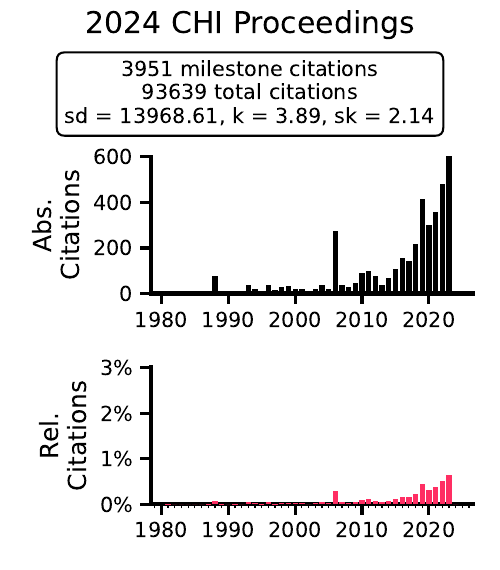}
  \\
  \caption{
  Absolute (top) and relative (bottom) number of citations from selected CHI proceedings to the 300 most cited milestones published in years prior to the given proceedings year.
  Note that for 2002 and 2004, the number of milestones is marginally lower.
  Standard deviation, kurtosis, and skewness values are listed in each plot.
  Looking at the panels, we can identify that while the absolute number of citations to prior milestones has increased, the relative number of citations to prior milestones has decreased significantly. In other terms, milestones have lost influence over time at CHI, relative to all other citations in a given proceedings year.%
  }%
  \label{fig:citationcurves}%
\end{figure*}%


The picture is more nuanced when
considering the evolution of citations at CHI (see \autoref{fig:citationcurves}).
  In \autoref{fig:citationcurves}, we can observe that the CHI community's citations have always been skewed (see \autoref{fig:citationcurves}), favoring more recent milestones over older ones.
An exception to this is the CHI community's preferential attachment to its super milestone which is clearly visible in recent proceeding years.
But while there have been small differences over the years, there is no clear trend toward citation distributions becoming more skewed in recent years.

There is, however, a notable shift between the absolute and relative number of citations to milestones in recent years.
In absolute terms, citations to the top-300 of CHI's milestones have increased, including citations to older milestones.
    The growth in the absolute number of citations in \autoref{fig:citationcurves} can be attributed to the extreme growth of references in CHI papers since 2016, as observed in prior work~\cite{oppenlaender-citations-HCI}.
In the 2024 proceedings alone, the CHI community made 3,951 citations to the prior top-300 milestones, compared to only 359 citations ten years earlier in the 2014 proceedings.
In relative terms, the 2024 number of citations to milestones pales in comparison to the total number of citations made in that proceedings year (93,639 citations).
    The top-300 prior milestones only account for a small fraction of citations in 2024 (4.2\%). 
This demonstrates that while the absolute number of citations to milestones has increased, the relative number of citations to milestones has declined.
    For comparison, the milestones published in the prior five years each received about 3\% of all citations in 2002 (top left in \autoref{fig:citationcurves}).
    In 2024, the relative number of citations to prior milestones is almost ten times lower (bottom right in \autoref{fig:citationcurves}).
    This highlights that in relative terms, older milestones are being cited less in the CHI community.


\begin{figure*}[!htb]%
\centering%
  \includegraphics[width=.85\linewidth]{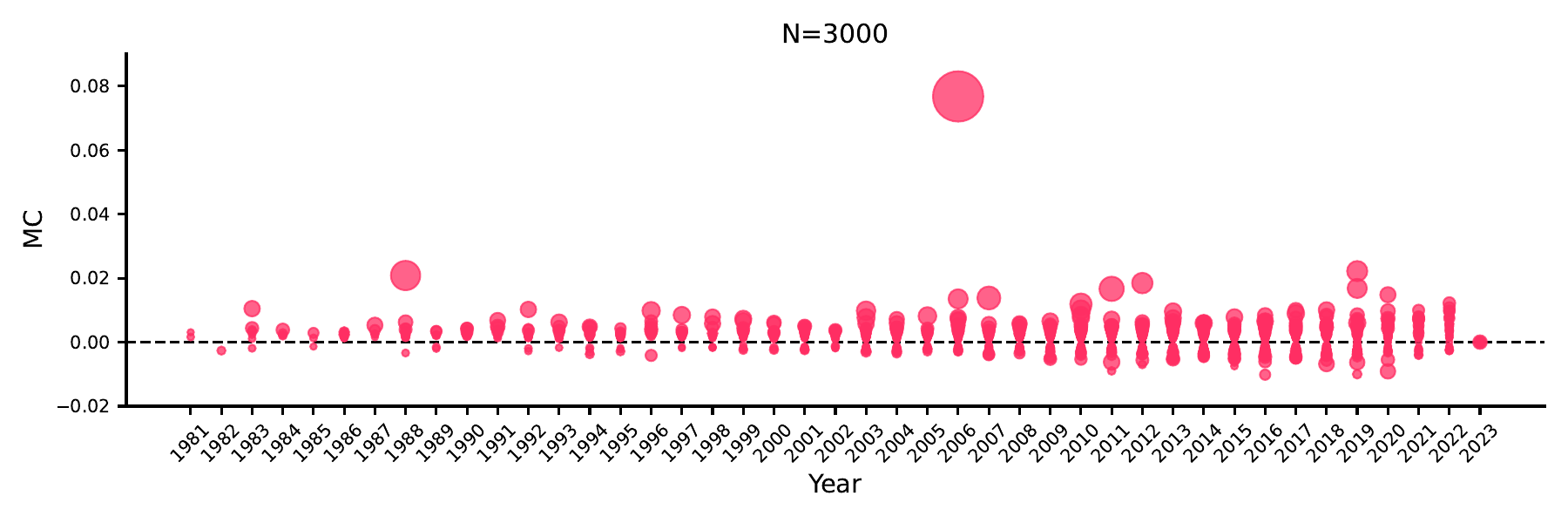}%
  \caption{Milestone coefficient (MC) for the 3,000~papers most cited by CHI authors. The year corresponds to the milestone's publication year, and the size of the bubble reflects the total number of citations to the milestone from CHI papers (over all CHI proceedings). Papers below the horizontal line tend to being forgotten by the CHI community over time, while papers above the line continue to grow in citations.
  The distance from the dashed horizontal line indicates the milestone's impact, with high-performing milestones having a greater positive distance from the line.
  The plot also demonstrates the exceptional performance of the super milestone published in 2006.
  }%
  \label{fig:milestonecoefficient}%
\end{figure*}%

We apply the milestone coefficient (MC) to 
the top-30 milestones (as listed in \autoref{tab:milestones}).
The MC in \autoref{tab:milestones} is negative only for one milestone (D3 \cite{D3}) which has seen a decline in citations from the CHI community in recent years. All other milestones in the top-30 list are still trending toward growth in citations.
To extend this analysis and derive insights on the CHI as a whole, we apply the milestone coefficient to the 3,000 papers most cited by CHI authors.
For these papers, the milestone coefficient ranges from -0.0102 to 0.0768.
\autoref{fig:milestonecoefficient} plots the MC for the top 3,000 most cited papers. Each bubble represents one paper in its year of publication. The size of the bubble represents the total number of citations to this milestone.
The exceptional super milestone is clearly visible in the year 2006, achieving an MC of 0.0768 (the highest MC in the dataset).
The second most cited milestone, the NASA TLX \cite{M2}, is visible in 1988.
From \autoref{fig:milestonecoefficient}, it can be derived that the number of milestones has grown over the years. This is unsurprising, given the growth of the CHI proceedings and the growth in citations \cite{oppenlaender-citations-HCI}.
However, the figure also demonstrates that many recent milestones struggle to receive the same attention from the CHI community as  past super milestones. Some of the recent milestones even trend towards a decline in citations, and the number of milestones that are being `forgotten' has increased in the last two decades.


\transition{%
Our analysis, so far, has been focused on citations and papers} as unit of analysis.
In the following section, we shift the focus to the milestone authors.%
%
%
%
\subsection{Does the CHI Community show a Preference for its Milestone Authors? (RQ4)}%
\label{sec:milestone-authors-attachment}%
\label{sec:milestone-authors}%
The CHI community is relatively large, with over a thousand authors attending the CHI Conference.
While attended by many first-timers, some authors at CHI are repeat authors \cite{1520340.1520364.pdf}.
The Matthew effect in science \cite{merton1968.pdf} describes the tendency of successful authors to be cited more often.
This accumulated advantage may also be present in the milestones at CHI, and we expect many milestone authors have authored more than one milestone.

\begin{figure}[!htb]%
\centering%
  \includegraphics[width=.95\linewidth]{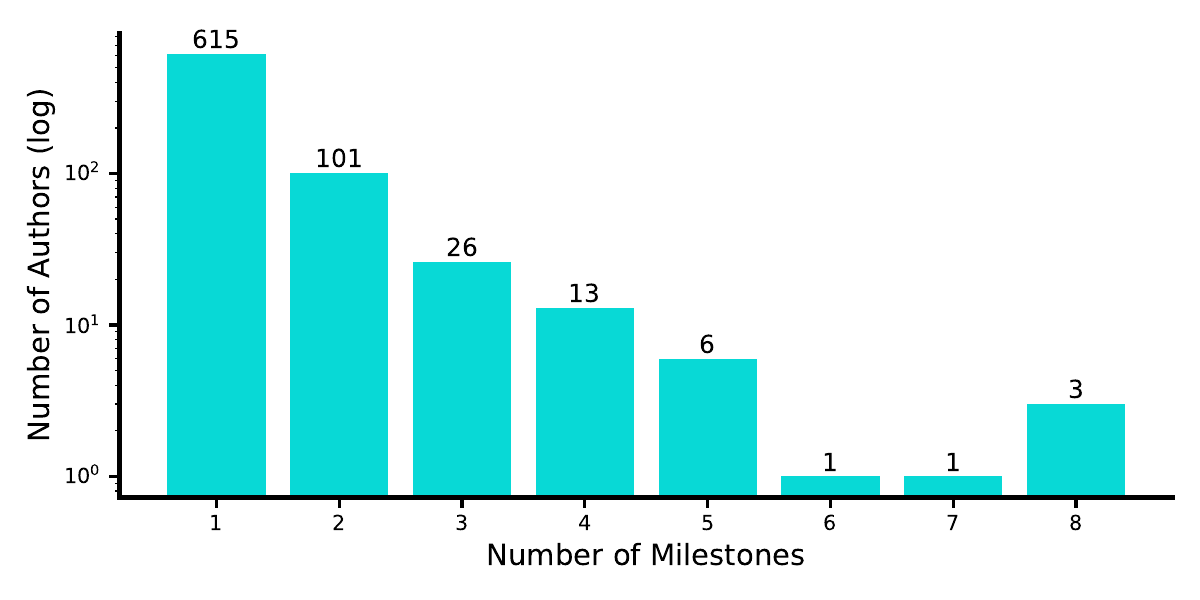}
  \caption{%
  A plot depicting the number of authors who have authored one or more papers among the CHI community's 300 most cited milestones.
  This includes articles published at CHI and external venues.
  While the majority of authors have only written one of the top-300 milestones, 
  about 20\% of the authors have (co-)authored multiple of the top-300 milestones.
  A logarithmic scale is used in this plot to make small values visible.%
  }%
  \label{fig:authors}
\end{figure}%

\begin{figure}[!htb]%
\centering%
  \includegraphics[width=\linewidth]{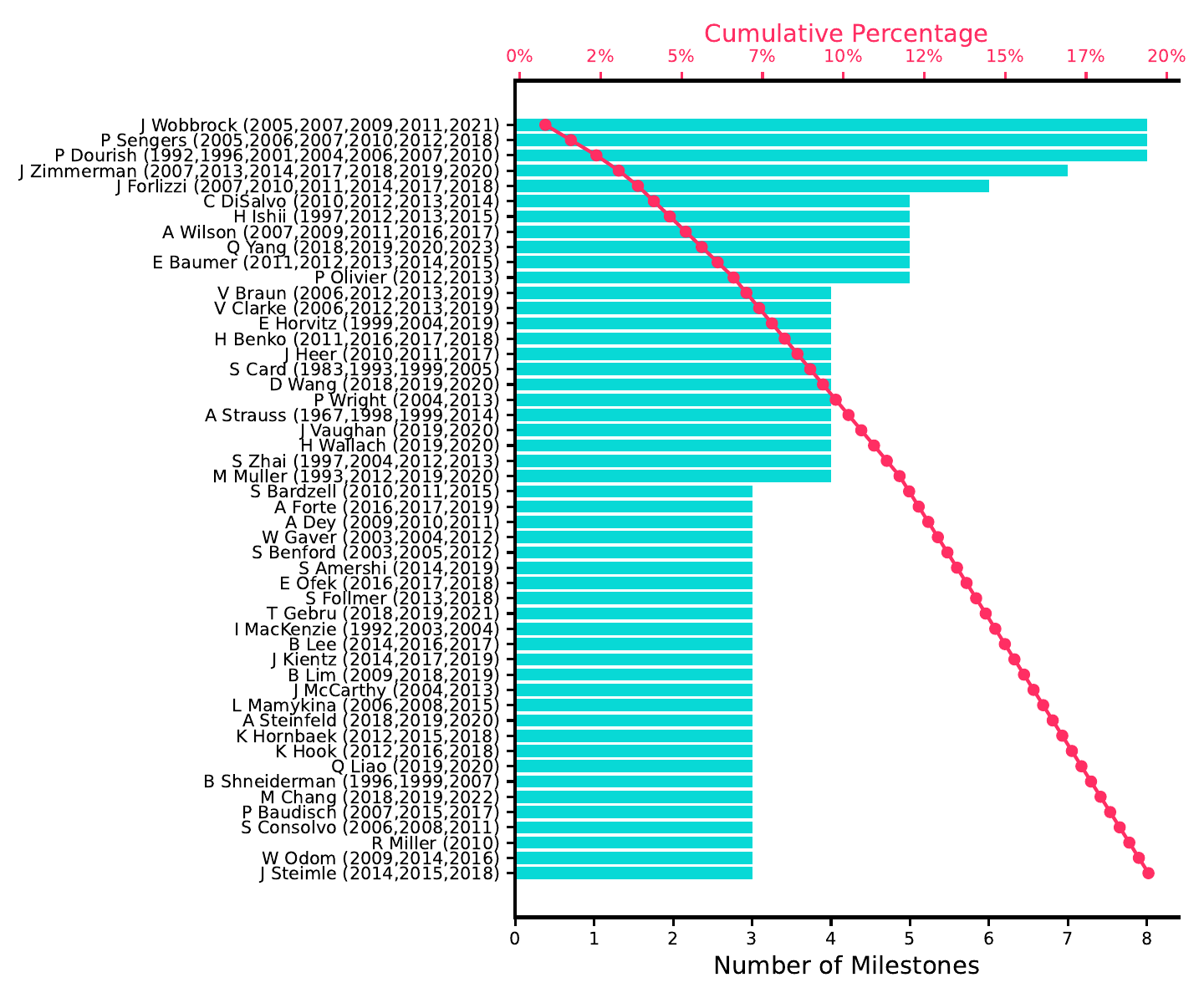}
  \caption{%
  Fifty authors have published three or more of the 300~milestones most cited in the CHI community (teal bars and lower x-axis).
  The 50~authors have \mbox{(co-)}authored  20\% of the 300 milestones (red line and top x-axis).
  The milestones publication years are listed in parentheses.
%
%
  }
  \label{fig:authornames}
\end{figure}

To investigate this, we identify the authors of the 300 milestones most cited in the CHI community (over all proceedings years)
    and count the number (see \autoref{fig:authors}) and cumulative percentage (see \autoref{fig:authornames}) of milestones these authors have (co-)authored.
We find that while the majority ($n=615$, 80.3\%) of the milestone authors have only authored one of the top-300 milestones, 151~authors (19.7\% of the top-300 milestone authors) have \mbox{(co-)}authored more than one of the top-300 milestones.
The 50~authors who have (co-)authored three or more of CHI's 300 most cited milestones are listed in \autoref{fig:authornames}.
    Chief among them are Phoebe Sengers, Paul Dourish, and Jacob O. Wobbrock who each have (co-)authored eight of the CHI community's 300 most cited milestones.
    This is followed by 
        John Zimmerman (7 milestones),
        Jodi L. Forlizzi (6 milestones),
        and others (see \autoref{fig:authornames}).
Together, the 50~authors 
have (co-)authored about 20\% of the 300~milestones most cited by the CHI community (see the secondary axis on top of \autoref{fig:authornames}).
%
\section{Discussion}%
\label{sec:discussion}%
The CHI community continues to grow in size and impact. 
In this paper, we presented meta-research on how the community operates as a scientific field.
Analyzing the way how we cite our milestones is not only an intellectually satisfying exercise, it also concretely explains what kind of work goes appreciated at large in the community.%
%
%
%
\subsection{The CHI Community's Milestones}%
\label{sec:discussion:milestones}%
What makes a milestone?
And what makes a paper stand the test of time?
Our research gives empirical indications of what is valued at CHI, as expressed through citations to milestone papers.
In our analysis, we found that milestones cited at CHI, and -- as a proxy, in HCI -- make mainly \textit{methodological} contributions.
These contributions deeply engage with the field's history and innovate over the state-of-the-art, resulting in a long-lasting impact in HCI.

This contrasts with some of the contribution types in HCI that are not represented in the list of top-cited milestones.
In particular, milestones are not being cited for contributing prototypes and systems, which is one very popular contribution type at CHI.
Spurred by the availability of large language models, it is now easier than ever to implement prototypes in HCI or write a paper about them~\cite{pang2025understandingllmificationchiunpacking}.
We can observe that some authors rush from one quickly implemented HCI study to the next \cite{d41586-023-02980-0.pdf}.
The methodological acceleration in HCI \cite{3699689.pdf} shifts our attention away from things that may have a long-lasting impact in HCI in favor of quick results.
Some researchers have raised concerns about the quick-win approach found in some HCI research, calling it concerning and detrimental to the field of HCI \cite{llmwrappers,3673861.pdf}. 
HCI has a strong case of ``shiny object syndrome'', where short-lasting projects are prioritized over time-persisting milestones.
We believe it is not likely that the increasingly popular ``thin-wrapper'' implementations with language models will become part of the set of milestone contributions in HCI.
It is important for us to not forget the past milestones and their contributions, as they inform our methodologies and values in HCI.

Milestones also do not include some other HCI research contribution types,
including
literature reviews, datasets, and opinions.
In particular, we note that literature reviews are a very young contribution type at CHI.
\citeauthor{oppenlaender-citations-HCI} found that literature reviews only became a topic at the CHI Conference after page restrictions were lifted in 2016, but have since grown in popularity  \cite{oppenlaender-citations-HCI}.
While it has also become more common to cite datasets and code repositories in CHI papers \cite{oppenlaender-citations-HCI}, datasets are not among the top milestones cited at CHI.
Opinions are, perhaps, the least known contribution type and, for this reason, not common at CHI.

Another type of contribution type that is not represented in the milestones are replications.
Much of the research in HCI is concerned with gathering primary data, which raises concerns for the replicability of HCI studies.
This ``replication crisis'' in science is a well-known problem \cite{533452a.pdf,3360311.pdf,pmed.0020124.pdf,s44271-023-00003-2.pdf}.
In HCI, several authors have studied replicability \cite{2556288.2557004.pdf,3170427.3188395.pdf}, but replication studies have not made it among CHI's top-cited milestones.
As \citeauthor{2559206.2559233.pdf} noted in their RepliCHI workshop summary at CHI 2014, HCI researchers ``have almost no drive and barely any reason to consider investigating the work of other HCI researchers'' \cite{2559206.2559233.pdf}.

This preference for certain contribution types at CHI has implications for authors. 
Scientists who want to contribute to HCI's long-lasting intellectual heritage should consider developing novel methods and frameworks that advance the field.
However, we acknowledge this is easier said than done, because these novel methods compete with existing methods, and novel methods may not receive attention from the community.
One example is \citeauthor{nexus}'s Nexus Analysis \cite{nexus}. This research method has first been presented in \citeyear{nexus}, but only a limited number of studies at CHI use this method (e.g., \cite{paper396.pdf} and \cite{3411764.3445139.pdf}).

These examples reflect how the CHI community prioritizes and balances its attention to past and present milestones.
Our work addresses a critical issue in the CHI community: the balance between innovation and the retention of foundational knowledge.
Older milestones may, over time, get pushed into the long-tail of the citation distribution where the milestones do not receive enough attention to rise to super milestone status.
This also suggests a shifting focus in the CHI community away from foundational works in favor of more recent contributions.
However, due to the extreme growth in the number of citations made in CHI papers, newer papers struggle to receive the collective attention of prior super milestones, and it is unclear whether newer papers will be able to rise to the level of fame of older super milestones.
It seems that most recent milestones, while relevant to some, fail to capture the CHI community’s collective attention compared to the success of past milestones.

The ``sweet spot'' of relevance for papers cited at CHI is 5--15 years back in time.
This  
range 
likely reflects a balance between novelty, relevance, accessibility, and consolidation of knowledge. 
Research within this range has typically undergone sufficient validation and follow-up, making it both trustworthy and broadly applicable, while remaining relevant to current advances.
This time frame allows foundational ideas to influence new work, often through citation cascades that amplify visibility and integration into broader academic and practical contexts.
Ideas from this period are also more likely to have been incorporated into curricula and practice, making them accessible and influential to a wide audience.
Similar patterns are seen in other disciplines, such as physics and medicine, where impactful research takes years to be validated, cited, and adopted into subsequent innovations.
Conversely, older work may become less cited as it is superseded by newer paradigms or rendered less accessible.
This highlights the dynamic interplay between innovation, validation, and knowledge dissemination in shaping research impact.



Another trend defining the dynamics of milestones is the growth in citations and publications at CHI.
As the CHI Conference evolved and scaled in diversity and size, milestone papers that concern the whole CHI community are less frequently occurring.
This development creates a dilemma for the CHI Conference: the conference has, arguably, become too large and diverse to have common denominators and shared research interests. What matters to one sub-community in CHI might be uninteresting to other sub-communities.

A community that does not celebrate its past milestones is a community that speedruns from one ephemeral project to the next.
If milestones are no longer valued and cited, and if the field fails to discover milestone papers in its publications, then, due to the growth of publications and the shifting attention of the field, it is becoming less and less likely every year for CHI to birth new milestones.
This is arguably concerning and endemic of the fractured nature of the CHI community in many sub-communities around different highly diverse topics \cite{2556288.2556969.pdf,oppenlaender2023mapping}.

These trends are indicative of a community at a crossroads between remembrance and forgetfulness,
a community that while honoring and valuing past contributions, is gravitating towards more recent work. 
Whether this is a problem or natural evolution is, of course, up for debate. 
These issues are known and have, of course, also been grappled with earlier, for instance through the \textit{CHI2030 Vision Task Force} (2018--2019) that sought to address the future of the CHI Conference.%
%
%
%
\subsection{Matthew Effect and Authoring at CHI}%
\label{sec:discussion:matthew}%
We demonstrated how certain prolific authors contribute a lot to the community milestones.
The 50 most highly-cited authors have \mbox{(co-)}authored about 20\% of the top-300 milestones cited at CHI, as depicted in \autoref{fig:authornames}.
There are numerous possible explanations for this concentration of prolific researchers.
Obviously, the work itself by these authors is of high-quality and constitutes advances worth remembering in new and upcoming research in the field of HCI. 
And some preferential attachment is normal in a research community.
On the other hand, some of the citation patterns might be explained through other forms of influence: large researcher networks, access to labs with plenty of funding and, thus, future researchers citing the past work from the same lab, etc. 
These are all factors that we should, in all fairness, recognize. It may be rather difficult to attain a milestone status for one's work, if it originates from an underdog position and from a relatively unknown laboratory. 
In any case, 
    we presented empirical evidence of a concentration of prolific milestone authorship in the CHI community.

What does this mean for authors at CHI?
The cumulative advantage and prominence of existing milestones may make it difficult for emerging authors to publish new milestones.
While it may arguably be easier for authors to contribute an impactful milestone in one of CHI's many sub-communities, authoring a milestone that concerns the whole community is hard, as new milestones may compete with the CHI community's attention and established milestones.
CHI's attachment to its one super milestone is the best example for this accumulated advantage.
    Is a grounded theory approach truly the best way to analyze (big) data in the twenty-first century?
    Some researchers have expressed their doubts and criticms about this \cite{deterding2018.pdf}.
The accumulated advantage and focused attention of the community on some highly performing milestones, we argue, may create barriers to entry for emerging milestones and their authors.

One could speculate on whether the CHI community is saturated with milestones, which would further contribute to barriers of entry.
    \citeauthor{M1}'s work, for instance, seems to receive fewer citations in recent years (relative to all citations in the CHI Proceedings), which could indicate that the CHI community has reached saturation with this 
    research method.
This may suggest that the CHI community is 
    at a 
    point where established milestones overshadow the emergence of new, impactful contributions.
In the following case study on CHI’s super milestone, we explore how \citeauthor{M1}'s influential milestone navigated these challenges, setting a precedent for how future milestones might overcome these potential barriers.%
\subsection{Case Study on CHI's Super Milestone}%
\label{sec:discussion:casestudy}%
Our case study on CHI's super milestone invites reflection on what makes a milestone enduring---beyond just time or relevance to the field.
    What criteria elevate a contribution to super milestone status and what does the CHI community implicitly value?
It is clear that Thematic Analysis provided a versatile method 
for qualitative data analysis.
This methodological guidance and research method proved to be a highly useful tool in the HCI researcher's toolbox.

However, \citeauthor{M1} have more recently acknowledged that many researchers apply their method not as it was originally intended~\cite{braun2019.pdf,MX}.
The authors themselves now advocate for the `fading' of their own milestone in favor of their more recent works  on Reflexive Thematic Analysis. 
In many cases, researchers only loosely apply the method as an umbrella for their (unstructured) qualitative investigations
or they apply different versions of the approach, as delineated by \citeauthor{braun2019.pdf} \cite{braun2019.pdf}.
\citeauthor{3613904.3642355.pdf} found the same issue in their review of autoethnographies, noting that ``existing   methods   are   often   executed   and   interpreted   with   much  
variety'' \cite{3613904.3642355.pdf}.
This speaks to how researchers often \mbox{(mis-)}appropriate methods and milestones.

\citeauthor{3544548.3581203.pdf} presented a scoping review on the use of Thematic Analysis in Healthcare HCI, noting that
the ``departures   from   what   is   advocated   as   quality  
[Thematic Analysis] practice'' constitute ``a change in research practices'' \cite{3544548.3581203.pdf}.
    In other words, the appropriation of the method by HCI researchers is testament to the innovation and continued evolving nature of the Thematic Analysis method. 
Like a spoken language that changes and evolves over time, a research method will be adapted and evolved by people who apply it.
But how can a research community refer to this evolved version of the method, if the method deviates significantly from its original version?
Clearly, many authors still attribute their evolved use of the method to the original milestone.
This is where articles such as \citeauthor{braun2019.pdf}'s later work \cite{braun2019.pdf} come in, because this article takes review of how the Thematic Analysis milestone was applied, 
and---crucially---still allows researchers to continue their use of the appropriated method, should they choose to.
    For instance, one can still use a 
    ``codebook'' approach to Thematic Analysis, and the correct milestone to cite for this approach is~\cite{braun2019.pdf}.%
%
%
%
%
%
\subsection{The Milestone Coefficient: A New Lens on Impact}%
\label{sec:discussion:coefficient}%
We introduced the Milestone Coefficient as a novel approach to quantifying the significance of contributions on a continuous scale.
The Milestone Coefficient is intended as a tool to identify milestones and reassess what constitutes a milestone in a research community, offering a fresh perspective that goes beyond traditional citation counts.
    More specifically, the Milestone Coefficient considers two factors. First, it considers the relative impact of a milestone compared with other papers in the field.
    Second, it considers the performance of a milestone against its own growth in citations.
    Scholarly papers will naturally accumulate citations over their lifetime, but this does not necessarily make them milestones. A milestone will show exceptional performance against its age-adjusted citations.

Many research indicators, such as the h-index, focus on citations.
Citations as a metric for evaluating papers are problematic for several reasons.
One reason is that milestones are often cited for the wrong reasons, as discussed in the previous section. Consequently, many milestone papers accumulate citations only for being a milestone.
Another reason is that citations have become extremely important to a researcher's career \cite{edwards-roy-2017-academic-research-in-the-21st-century-maintaining-scientific-integrity-in-a-climate-of-perverse.pdf}.
A recent survey of over 30,000 faculty members of the 10~highest ranked universities globally found that Google Scholar is the most popular source of information to evaluate a researcher's career \cite{2402.04607.pdf}.
The same study demonstrated that citation counts in Google Scholar can be manipulated, which has also been demonstrated by other studies (e.g., \cite{Ike_Antkare}).
Citations, therefore, are a shallow proxy for research success and can be manipulated.

The Milestone Coefficient is by no means intended as a tool for research assessment or a substitute to the h-index.
The MC, as yet another bibliometric metric, will not help with addressing the current problems with bibliometrics and academia.
    Further, only very few authors will---by definition---become milestone authors.
    For this reason, the Milestone Coefficient would be ineffective for the purpose of evaluating the career progress of most researchers.
Rather, our intention with the Milestone Coefficient is
to provide a tool and quantitative means for identifying milestones, discussing, and comparing the performance of highly-cited milestones, going beyond a discrete typology of milestones.
We hope this coefficient will provoke a discussion on
the works that have had
the greatest impact at CHI and about what we value in the HCI research community.%
%
%
%
\subsection{Limitations and Future Work}%
\label{sec:limitations}%
We acknowledge limitations to the validity of our research.
%
First, while the ACM CHI Conference is a leading venue in HCI research and, thus, could be viewed as a suitable proxy for investigations on the wider field of HCI, our investigation is limited to our subject of study: the CHI community.
It is our intention to understand the citation behavior within this particular community, and how the community has evolved over time.
Future work could expand to other research communities, providing valuable comparison points.

Second, our work examines how the CHI community cites its milestones.
It is, however, possible that the CHI community favors derivative works over the original milestones.
    Metaphorically, authors of derivative works stand on the shoulders of giants, but by becoming giants themselves, they may crush what is underneath them.
One example of this is Fitts' Law \cite{fitts}.
    Fitts' highly influential work was published in \citeyear{fitts} and inspired much research in HCI.
    However, Fitts' success was shared by other authors who presented studies on Fitts Law.
While we do not find strong evidence for other instances of this in our research, future work could investigate how newer milestones and derivative works displace attention to older milestones.

Third, we acknowledge that our method of identifying milestone papers and authors is imperfect.
    We uniquely identify papers by their title and publication year. We use regular expressions to extract the paper titles. While we use normalization (such as lower casing and removal of punctuation and white spaces), our simple string matching approach is limited in uniquely identifying papers. References in CHI papers are noisy and may contain human mistakes, such as errors and typos. Our method cannot address these human errors.
In a similar vein, author name disambiguation is 
a common problem in bibliometric analyses
that is impossible to solve completely~\cite{1518701.1518810.pdf,1520340.1520364.pdf}.
    For instance, authors may use different versions of their name (e.g., see the first author in \cite{SIMO} and \cite{SIMOJOHANNES}), or they may legally change their name entirely.
While we did our best to develop a set of regular expressions to capture different notations of author names, human errors can not be fully accounted for.

Fourth, 
the growth in the number of papers and the increase in the number of references included in CHI papers may contribute to gradual smoothing of the forgetting curves over time.
This would not be an issue if changes were proportional (such as a proportional increase in the number of references and papers).
However, as described by \citeauthor{oppenlaender-citations-HCI}, there was a shift in citation behavior in 2016 that constitutes a non-linearity in the CHI community's trajectory~\cite{oppenlaender-citations-HCI}.
The forgetting curves should be interpreted with this 
non-linearity
in mind.

Last, our methodology focuses on total citations over the entire ACM CHI Conference. This methodology is not likely to identify milestones that have been popular for a very short period of time but since have 
faded completely from CHI's collective memory.
This would not allow the milestones to have accrued enough citations to be noticeable in our methodology.
We leave
    this
and investigations of milestone impact in CHI's sub-communities to future work.%
%
%
%
%
\section{Conclusion}%
\label{sec:conclusion}%
We presented an analysis of how the CHI community is, over time, shifting its attention away from 
its past milestone papers.
Forgetting curves provide graphical evidence
that the relative attention of the CHI community to its milestone publications is decreasing over time.
The attention span of the CHI community narrowly rests on milestone papers published in about a ten year window.
While this is a testament for the dynamism and renewal of research at CHI and in the field of HCI, it raises concerns about the value, sustainability, and meaningfulness of CHI research.
The HCI field's key contributions consist of methodological and conceptual contributions that move the field forward.
However, it is unclear whether recent milestones can achieve the fame of past milestones in HCI.

Our findings and the Milestone Coefficient offer practical insights for researchers and policymakers in HCI.
We hope our work will provide insights into the dynamics of research recognition and longevity in HCI, provoke a critical reflection in the CHI community on its fundamental values, and foster discussion on sustainable strategies forward.%
\begin{acks}
This research was partially funded by the Strategic Research Council (SRC), established within the Academy of Finland (Grants 356128, 335625, 335729), and Academy Research Fellow funding by Academy of Finland (Grants 356128, 349637 and 353790).
\end{acks}

\bibliographystyle{ACM-Reference-Format}
\bibliography{paper}


\end{document}